\documentclass[%
 reprint,
 amsmath,amssymb,
 aps,
]{revtex4-1}

\usepackage{graphicx}% Include figure files
\usepackage{dcolumn}% Align table columns on decimal point
\usepackage{bm}% bold math

	\newcommand{\ket}[1]{\left| #1 \right\rangle}

\begin{document}

%\preprint{APS/123-QED}

\title{Phonon number sensitive electromechanics}% Force line breaks with \\

\author{J. J. Viennot}
\email{viennot@jila.colorado.edu}
\author{X. Ma}
\author{K. W. Lehnert}
%\email{konrad.lehnert@jila.colorado.edu}

\affiliation{ JILA, National Institute of Standards and Technology and the University of Colorado, Boulder, Colorado 80309, USA \\
and Department of Physics, University of Colorado, Boulder, Colorado 80309, USA
}

%\author{Charlie Author}
% \homepage{http://www.Second.institution.edu/~Charlie.Author}
%\affiliation{
% Second institution and/or address\\
% This line break forced% with \\
%}%
%\affiliation{
% Third institution, the second for Charlie Author
%}%
%\author{Delta Author}
%\affiliation{%
% Authors' institution and/or address\\
% This line break forced with \textbackslash\textbackslash
%}%

\date{\today}% It is always \today, today,
             %  but any date may be explicitly specified

\begin{abstract}

We use the strong intrinsic non-linearity of a microwave superconducting qubit with a 4 GHz transition frequency to directly detect and control the energy of a micro-mechanical oscillator vibrating at 25 MHz. The qubit and the oscillator are coupled electrostatically at a rate of approximately $2\pi\times$22 MHz. In this far off-resonant regime, the qubit frequency is shifted by 0.52 MHz per oscillator phonon, or about 14\% of the 3.7 MHz qubit linewidth. The qubit behaves as a vibrational energy detector and from its lineshape we extract the phonon number distribution of the oscillator. We manipulate this distribution by driving number state sensitive sideband transitions and creating profoundly non-thermal states. Finally, by driving the lower frequency sideband transition, we cool the oscillator and increase its ground state population up to 0.48$\pm$0.13, close to a factor of 8 above its value at thermal equilibrium. These results demonstrate a new class of electromechanics experiments that are a promising strategy for quantum non-demolition measurements and non-classical state preparation. 

\end{abstract}

\pacs{Valid PACS appear here}% PACS, the Physics and Astronomy
                             % Classification Scheme.
%\keywords{Suggested keywords}%Use showkeys class option if keyword
                              %display desired
\maketitle

%\tableofcontents

The ability to bring man-made acoustical or mechanical structures into the quantum regime has been demonstrated in a variety of devices, from micro-mechanical oscillators in opto- and electro-mechanics experiments \cite{Teufel2011,Chan2011}, to acoustic resonators in circuit Quantum Electrodynamics (cQED) experiments \cite{O'Connell2010}. Mechanical oscillators are generally very linear harmonic oscillators at the quantum scale and to achieve arbitrary quantum control, one needs an extrinsic non-linearity \cite{Braunstein2005}. Performing non-linear detection is also a way to enable quantum non-demolition measurement by measuring energy instead of position or momentum \cite{Caves1980}. One strategy is to use the Josephson junction used in superconducting microwave circuits. It provides a dissipationless strong non-linearity and has enabled the demonstration of landmark results in quantum science from the preparation of arbitrary quantum states of microwave light \cite{Hofheinz2009,Vlastakis2013} to the demonstration of early-stage quantum computers \cite{Kelly2015,Kandala2017}. By using piezoelectric materials, resonant coupling between superconducting qubits and high frequency (GHz) acoustic wave resonators has been demonstrated \cite{O'Connell2010,Chu2017}. %In this regime, a quantum of energy can be coherently swapped between the qubit and the motional degree of freedom if the coupling rate exceeds the decoherence rates. 

This resonant approach is however restricted to a small class of acoustic oscillators and loses many of the advantages of the micro-mechanical oscillators used in opto- and electro-mechanics experiments \cite{Aspelmeyer2014}. In these experiments, a wide variety of techniques have been developed and have made these mass-on-a-spring-like oscillators very versatile. They can be used to interface otherwise incompatile quantum systems such as superconducting circuits and optical light \cite{Andrews2014}, they are extraordinarily sensitive detectors of force and strain \cite{Abbott2016Short,Moser2013} and they can be engineered to have extremely long lifetimes \cite{Ghadimi2018}. However these low frequency mechanical oscillators have proven to be more challenging to couple to superconducting qubits. One strategy is to use a linear cavity to transfer nonclassical microwave fields created by a qubit to a mechanical oscillator by using the radiation pressure interaction \cite{Lecocq2015,Reed2017}. This approach has to battle the incompatibility of large microwave pump powers with qubits as well as the loss during the state propagation or transfer. Low frequency mechanical oscillators have also been directly coupled to qubits \cite{LaHaye2009,Pirkkalainen2013a}, but so far the interaction strengths have been too weak to achieve control or detection of motion at the scale of few phonons. 

\begin{figure} [ht] \centering
\includegraphics[width=1.0\columnwidth]{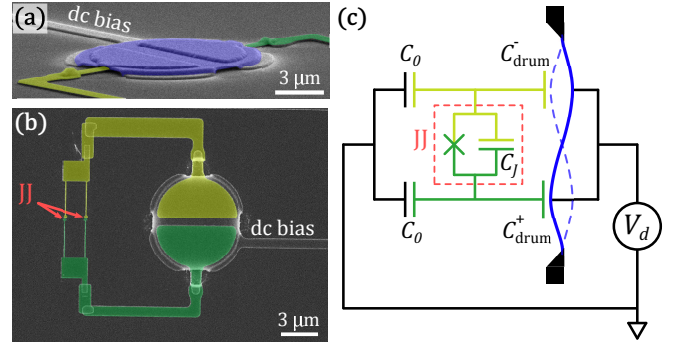}
\caption{(a) False-colored scanning electron micrograph (at an angle) of the micro-mechanical oscillator (blue) suspended above two electrodes and forming a vacuum gap capacitor. (b) Top view of the device showing the superconducting qubit electrodes (yellow and green), shunted by two Josephson junctions (JJ) in parallel. The dc bias line imposes a voltage $V_d$ onto the oscillator plate. (c) Equivalent circuit of the device. }
\label{Fig1}
\end{figure}

In this Letter, we directly couple a superconducting qubit to a mechanical oscillator, achieving an ultrastrong interaction of $g_m \approx 2\pi\times 22$ MHz, comparable to the oscillator's resonance frequency $\omega_m \approx 2 \pi\times 25$ MHz. Similar to quadratic optomechanics proposals \cite{Thompson2008}, we detect the energy of the oscillator instead of its position. More precisely, a mechanical ac-Stark effect shifts the qubit frequency by 0.52 MHz per oscillator phonon, or about 14\% of the 3.7 MHz qubit linewidth. The qubit lineshape therefore encodes the phonon number statistics, which we extract using a Bayesian-based algorithm. The qubit-oscillator system also exhibits blue and red sideband transitions, analogous to those found in optomechanics and trapped ions systems \cite{Aspelmeyer2014,Leibfried2003}, at the sum (blue) and difference (red) of frequencies. In contrast to optomechanics, the qubit non-linearity makes these sideband transitions number state dependent. Using this property, we demonstrate control of populations in the Fock space with a resolution of about 7 quanta. By driving the lower frequency sideband transition, we cool the oscillator and increase its ground state population up to 0.48$\pm$0.13, close to a factor of 8 above its value at thermal equilibrium.

Because our mechanical oscillator frequency is so low compared to that of a qubit at a few GHz, we require a qubit that couples statically to the oscillator. We achieve this interaction by forcing electrostatic charge onto the oscillator and using a Cooper-pair box qubit \cite{Bouchiat1998,Makhlin2001}, which unlike a transmon, is sensitive to charge at low frequency \cite{Koch2007}. Our device is presented in Fig. 1(a) and (b), and it can be mapped on the circuit diagram of Fig. 1(c). The mechanical oscillator is an aluminum drumhead, similar to those used in previous electromechanics experiments \cite{Teufel2011}. It is suspended over two separate aluminum electrodes and realizes a mechanically compliant capacitor with each electrode. When the drum vibrates in its first harmonic, the regions of the drumhead located above the two bottom electrodes move with opposite phases as depicted in Fig. 1(c). The two bottom electrodes are connected through two Josephson junctions in parallel, which hybridize the charge states of the two islands to form a flux tunable charge qubit \cite{Bouchiat1998,Makhlin2001}. We operate this qubit at the charge degeneracy point. The device is embedded in a far detuned co-planar waveguide resonator such that the qubit can be readout and coherently controlled using standard cQED techniques \cite{Wallraff2004,NoteSM}.

Our ability to apply a large dc bias on the drum is essential to the working principle of the device.  When such a voltage is applied, a static charge accumulates on the capacitors plates $C_{\textrm{drum}}^{\pm}$ and this charge is forced to move along with the mechanical motion. The motion of this charge is equivalent to an ac voltage applied differentially over the qubit electrodes and creates a Josephson current going through the junctions at mechanical frequency. Mechanical motion is thus transversely coupled to the qubit transition, realizing a Rabi Hamiltonian with a coupling strength $g_m \approx 2 \pi \times 3.7$ MHz/V. Most of the data we present was obtained with $V_d = 6$ V. 

\begin{figure} [ht] \centering
\includegraphics[width=1.0\columnwidth]{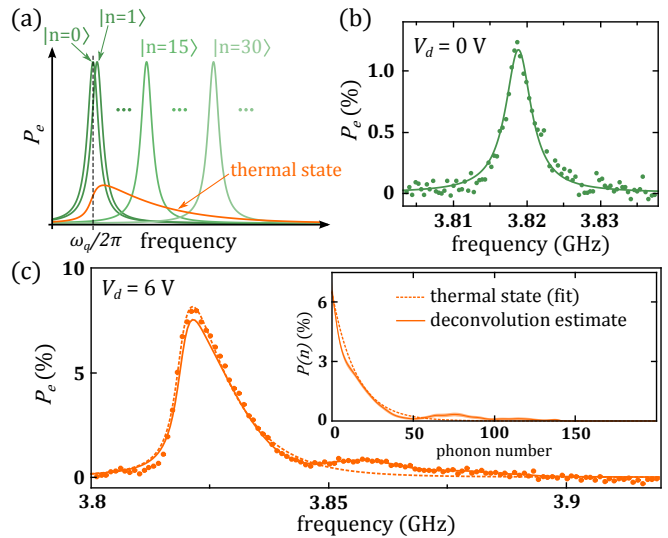}
\caption{(a) Principle of the ac-Stark shift. The bare qubit resonance is shifted by a fraction of its linewidth for each number state. For a mechanical thermal state, the dressed qubit lineshape is the sum over all number states weighted by their population. (b) Low power spectroscopy of the qubit decoupled from the mechanical oscillator, at $V_d =$ 0 V. The solid line is a Lorentzian fit indicating an intrinsic qubit linewidth of about 3.7 MHz. (c) Spectroscopy of the qubit coupled to the mechanical oscillator (at thermal equilibrium). Inset : phonon populations extracted with a fit assuming a thermal distribution (dashed) or with a Bayesian-based deconvolution algorithm (full line). }
\label{Fig2}
\end{figure}

\begin{figure*}[ht] \centering
\includegraphics[width=1.0\textwidth]{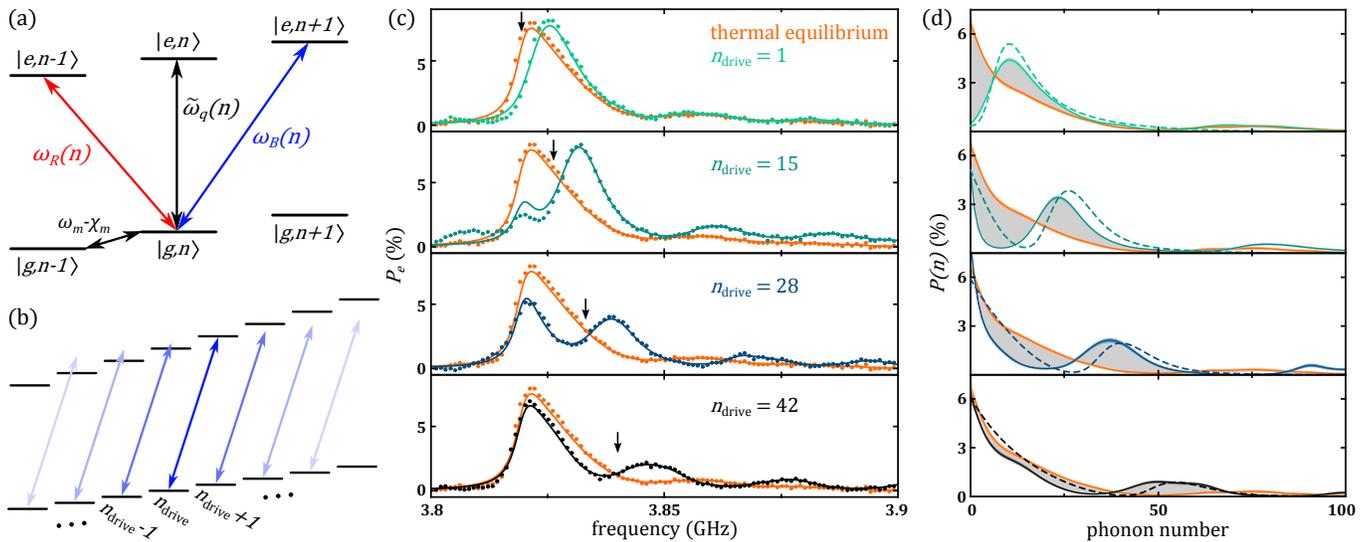}
\caption{(a) Energy level diagram for ground and excited states of the qubit ($g$ and $e$), dressed by phonon numbers $n-1$, $n$ and $n+1$, showing the number-sensitive qubit transition $\tilde{\omega}_q(n)$ as well as the number-sensitive blue and red sideband transitions $\omega_{B(R)}(n)$. (b) For a blue sideband drive centered around $n_{\textrm{drive}}$, blue sideband transitions at neighboring number states are also driven, at a smaller rate. (c) Qubit spectroscopy after a blue sideband drive centered at a few different $n_{\textrm{drive}}$. Dots are raw data, full lines are qubit lineshapes expected from the reconstructed phonon number distribution. Vertical black arrows indicate the position of $n_{\textrm{drive}}$ mapped onto the spectroscopic frequency axis (that is $\omega = \omega_q + 2 \chi_m n_{\textrm{drive}}$). (d) Reconstructed experimental phonon populations (full lines) and master equation simulation (dashed lines) based on independently measured parameters. Confidence intervals on the reconstruction (lighter shade) are obtain from a non parametric bootstrap \cite{NoteSM}. Gray areas show where populations have been moved by the sideband drive. }
\label{Fig3}
\end{figure*}

At zero bias voltage, the motion of the drum only modulates the capacitances of the qubit electrodes and the qubit-mechanics coupling is negligible. We can use this fact to characterize the bare qubit. Figure 2(b) shows a qubit spectroscopy at $V_d = $ 0 V, obtained by measuring the qubit excited state probability $P_e$ as a function of the frequency of a weak microwave drive. Fitting this to a Lorentzian, we obtain a FWHM of about $\Gamma_2^* \approx 2 \pi \times 3.7$ MHz, consistent with coherent control data in the time domain \cite{NoteSM}.  

When a 6 V dc bias is applied on the drum, the qubit and the oscillator have a very large coupling strength ($g_m / \omega_m \approx 0.9$). Nevertheless, this device is in a regime where the mechanical frequency is more than two orders of magnitude smaller than the qubit frequency. In this limit, the two systems do not exchange energy spontanously but instead they shift each other's resonance frequency according to the effective Hamiltonian \cite{Blais2004,Wallraff2004,Schuster2005,Beaudoin2011} 
\begin{equation} \label{eq:Hamiltoninan}
H/\hbar = \omega_m a^\dag a + \frac{1}{2} \left( \omega_q + 2\chi_m a^\dag a \right) \sigma_z
\end{equation}
where $a$ is the phonon annihilation operator, $\sigma_z$ is the qubit Pauli operator, $\omega_q$ is the Lamb-shifted qubit frequency and $\chi_m \approx 2 g_m^2 / \omega_q$ includes the Bloch-Siegert shift \cite{Bloch1940,Zueco2009,Beaudoin2011}. We determine $2 \chi_m \approx 2 \pi \times$0.52 MHz by measuring how the mechanical oscillator frequency is dispersively shifted by the qubit in its ground state \cite{NoteSM}. Equation 1 shows how the qubit transition frequency is dressed by each phonon in the oscillator, $\tilde{\omega}_q(n) = \omega_q+2\chi_m n$. This effect is sketched in Fig. 2(a). A figure of merit of our device is given by the ratio between the single phonon Stark shift and the qubit FWHM, $2 \chi_m / \Gamma_2^* \approx 0.14$. In terms of resolution, this means that the sum of two phonon number states different by 7 or more will yield a qubit spectrum with two resolved peaks. As shown below, performing fits or deconvolutions allows us to go beyond this limit. 

In order to determine the phonon number distribution from the ac-Stark shift, we make an approximation that treats the qubit lineshape dressed by the mechanical motion as \cite{Gambetta2006}
\begin{equation} \label{eq:Convolution}
P_e^{\textrm{dressed}}(\omega) = \sum_n P(n) P_e^{\textrm{bare}}(\omega-2\chi_m n)
\end{equation}
where $P_e^{\textrm{bare}}(\omega)$ is the response of the bare qubit to a spectroscopic drive at the frequency $\omega/2\pi$. The validity of this approximation depends on three conditions which are well satified for phonon numbers up $\approx$ 50 \cite{Clerk2007a,NoteSM}. This holds us from being quantitative for populations at larger phonon numbers with equation 2, but as discussed below, it allows us to make qualitative statements about arbitrary distributions. 

   % First, the system must be in the weak damping limit $\gamma_m \ll \chi_m$, with $\gamma_m$ the mechanical damping rate. This condition ensures weak back-action \cite{Clerk2007a}, so that the height of each underlying peak is $P(n)$. Second, the width of each underlying peak $n$ (approximately given by $\Gamma_2^* + 2 n n_{th} \gamma_m$ \cite{Clerk2007a}, $n_{th}$ being the average thermal occupancy) is dominated by the bare qubit linewidth $\Gamma_2^*$. Third, the number of phonons is small compared to $\omega_m / 2 \chi_m$ to avoid unfortunate degeneracy between dressed qubit transitions and second order sideband transitions \cite{NoteSM}. 
   % We measure $\gamma_m \approx 2 \pi \times$ 94 Hz \cite{NoteSM} and $n_{th} \approx 15$. This ensures the first condition to be very well satisfied. The second condition is valid until about 300 phonons and corrections can easily be added to equation 2 to account for broadening \cite{NoteSM}. The third condition is only valid up to about 50 phonons, but modifying equation 2 to account for large phonon numbers is a more challenging task which is beyond the scope of this paper. This holds us from being quantitative for populations at large phonon numbers, but as discussed below, equation 2 allows us to make qualitative statements about arbitrary distributions.

The measured spectroscopy of the qubit dressed by the mechanical oscillator at thermal equilibrium with our dilution refrigerator is shown in Fig. 2(c). The asymmetry in the lineshape reflects the thermal distribution of the oscillator, with a tail going to high Fock numbers. Assuming that the mechanical oscillator number state distribution is that of a thermal state, $P(n)=n_{th}^n / \left( n_{th}+1 \right)^{n+1}$, we can fit this lineshape using equation 2 with $n_{th}$ as a fit parameter and we extract $n_{th} \approx 15$ (about 18 mK) \cite{NoteSM}. This procedure only works for mechanical states for which we know the functional form of $P(n)$ (thermal states, coherent states, displaced thermal states, etc). In order to be more general, we use an adapted version of the Bayesian-based Lucy-Richardson algorithm to invert equation 2 \cite{Richardson1972,NoteSM}. The extracted $P(n)$ distribution is shown in Fig. 2(c), along with the distribution obtained assuming a thermal state. The reconstructed populations are plotted along with confidence intervals obtained from a non-parametric bootstrap \cite{Efron1994}. The small bump in the data, just above 3.85 GHz, is a manifestation of a deviation from the small number of phonons approximation, which makes equation 2 inexact.

We now use the qubit to control the energy distribution of the mechanical oscillator by driving sideband transitions. Similar to opto- or electro-mechanics systems, we can drive a red or blue sideband transition. As depicted in Fig. 3(a), a red sideband transition excites the qubit while removing a phonon from the oscillator. Conversely a blue sideband transition excites the qubit and adds a phonon to the oscillator. The crucial difference from that of conventional linear opto-mechanics is the number-state dependence of these transitions. The blue (red) transition frequencies are given by :
\begin{equation}
\omega_{B(R)}(n) = \omega_q \pm \omega_m  + 2 \chi_m \left( n \pm \frac{1}{2} \right)
\end{equation}
Thus driving a blue sideband transition at frequency $\omega_{B}\left( n_{\textrm{drive}} \right)$ only drives a few transitions at neighboring number states. The characteristic number of transitions being driven around $n_{\textrm{drive}}$ is given by $\Gamma_2^*/ 2 \chi_m $ \cite{NoteSM}, that is on the order of 7 transitions. Because these are two-photon transitions, we drive these sidebands with two tones: a lower frequency (260 MHz) dither applied to the Cooper-pair box gate, and a microwave tone detuned from the transition by the dither frequency \cite{NoteSM}, as proposed in Ref. \cite{Blais2007}.

%As depicted in Fig. 3(b), because our qubit linewidth exceeds the single phonon Stark shift $2\chi_m$, driving a blue sideband at frequency $\omega_{B}\left( n_{\textrm{drive}} \right)$ also drives the transitions at neighboring number states. The number of transitions being driven around $n_{\textrm{drive}}$ is given by the figure of merit of our device, $2 \chi_m /\Gamma_2^*$ \cite{NoteSM}, that is on the order of 7 transitions. At charge degeneracy, the symmetry of the system Hamiltonian implies that these transitions absorb two photons from the drive. Instead of driving intensely at half the transition frequency, we apply two tones: a lower frequency (260 MHz) dither to the Cooper-pair box gate, and a microwave tone detuned from the transition by the dither frequency \cite{NoteSM}, as proposed in Ref. \cite{Blais2007}. The sideband transitions Rabi rates are then linear in the microwave tone amplitude.

Figure 3(c) shows qubit spectroscopies taken after a microwave pulse at a few different blue sideband frequencies, corresponding to transitions ranging from $n_{\textrm{drive}}\approx$ 1 to 42. When driving close to the ground state ($n_{\textrm{drive}}\approx$ 1), the dressed qubit resonance is essentially shifted to slightly higher frequency. However, driving at higher numbers qualitatively changes the qubit lineshape, showing a first peak around the bare qubit frequency and a separate peak at higher frequency. This lineshape reveals how the phonon populations are moved into distinct regions of the Fock space. The associated reconstructed phonon distributions $P(n)$ are given in Fig. 3(d). As highlighted by the shaded areas, these distribution show how our blue sideband pulse takes phonon populations around $n_{\textrm{drive}}$ (or below), and transfers them to higher numbers. We can compare the reconstructed experimental phonon distributions with a semi-classical master equation simulation (dashed lines in Fig. 3(d)). In these simulations, the qubit decoherence rates, the mechanical damping rate, the mechanical bath temperature and the sideband rates are determined from independent measurements \cite{NoteSM}. We attribute the difference between experiment and simulation mainly to deviations from equation 2 that yield inaccurate reconstructions. Nevertheless, the qualitative agreement between simulation and experimental data demonstrates how the qubit can be used to control the phonon population in the drum with a resolution of a few phonons and up to a relatively large number of phonons. 

\begin{figure} [t] \centering
\includegraphics[width=1.0\columnwidth]{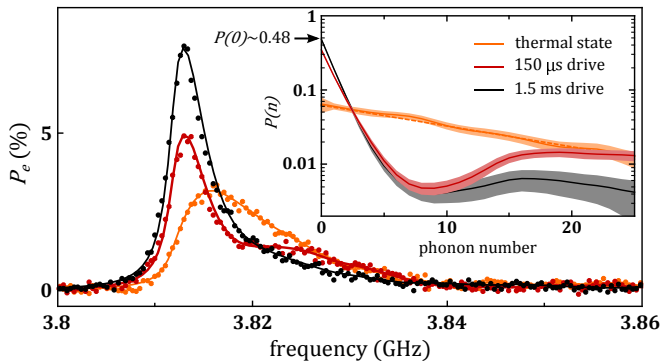}
\caption{ Spectroscopy of the qubit after a red sideband drive of 150 $\mu$s (dark red) and 1.5 ms (black), centered around $n_{\textrm{drive}} \approx 8$. Dots are raw data, full lines are qubit lineshapes expected from the reconstructed phonon number distribution. Inset : reconstructed experimental phonon populations (log scale, with bootstrap confidence intervals in lighter shades). The dotted line is the distribution of a fit to a thermal state. The horizontal black arrow indicates the ground state population of about 0.48 (see text).}
\label{Fig4}
\end{figure}

Finally, Fig. 4 shows how we use the qubit to cool thermal motion down with a red sideband drive. Cooling macroscopic mechanical motion with a superconducting qubit has been extensively investigated theoretically \cite{Martin2004,Zhang2005,Hauss2008,Jaehne2008,Rabl2010}, but to our knowledge this is the first experimental demonstration of such a scheme. After a 150 $\mu$s red sideband pulse at $n_{\textrm{drive}} \approx 8$, the population around $n=8$ is transfered to Fock numbers $n\leq 2$ (red data in Fig. 4). Leaving the sideband drive on for a longer time, comparable to the mechanical damping time of about 1.7 ms, population at higher energy slowly decays down into the range where the sideband drive is effective (black data). This further increases the populations around the mechanical ground state to reach $P(0)\approx 0.48 \pm 0.13$. The uncertainty on $P(0)$ is here dominated by our uncertainty in the bare qubit frequency \cite{NoteSM} (uncertainty bands in Fig. 4 do not include this systematic effect). After this long pulse, the narrowed qubit lineshape is a direct signature of the decreased phonon variance associated with lower mechanical temperature. For longer duration red-sideband pulses, the drive begins to trivially heat the oscillator rather than cool it. Nevertheless, the demonstrated performance should be sufficient to prepare sub-Poissonian states at large average phonon number. 

%We have demonstrated the coupling of a superconducting charge qubit to a low frequency micro-mechanical oscillator. We have shown how the qubit is a non-linear detector sensitive to energy. Using number-sensitive sideband transitions, we have used the qubit to control phonon populations and to cool down mechanical motion. 

Looking forward, a natural next step would be to increase $\chi_m$ by increasing $V_d$. The maximum voltage we could apply in this study (6 V) was limited by our ability to readout the qubit through its dispersive coupling to the microwave resonator. For reasons we do not understand the readout contrast diminished and became bystable with increasing voltage. Understanding and solving this problem would allow us to turn up $V_d$, in principle up to 21 V (limited by electrostatic instability). The single phonon Stark shift $2\chi_m$ would then be approximately $2\pi\times$10 MHz, exceeding the bare qubit linewidth and reaching the strong dispersive limit \cite{Schuster2007a}. In addition, the ultra-strong qubit-mechanics interaction demonstrated here could also be combined with the microwave cavity to enter a rich three-body interaction regime \cite{Abdi2015}. This could be used to prepare non-classical states such as mechanical cat states and tripartite entangled states involving the microwave cavity, the qubit and the mechanical oscillator. \\

We acknowledge Shlomi Kotler and Karl Mayer for enlightening discussions. We gratefully acknowledge Ray Simmonds and Florent Lecocq for their help with the fabrication of the device. We acknowledge fruitful discussions with Felix Beaudoin, Joe Aumentado, William Kindel, Michael Schroer, Lin Tian, Xian Wu, David Pappas, John Teufel and Robert Lewis-Swan. We thank Maxime Malnou and Daniel Polken for providing us with a Josephson parametric amplifier. We acknowledge funding from National Science Foundation (NSF) under Grant Number 1125844 and from the Gordon and Betty Moore Foundation. %\textbf{Author contributions:} J.J.V. and X.M. designed and fabricated the device, performed the experiment, analyzed the data, and performed the numerical simulations, all under the supervision of K.W.L.; J.J.V. wrote the manuscript with inputs from X.M. and K.W.L. \textbf{Competing interests:} The authors declare no competing interests. \textbf{Data and materials availability:} All data needed to evaluate the conclusions in the paper are present in the paper and the supplementary materials.

\pagebreak
\newpage

\begin{widetext}

\begin{center}
\large{\textbf{Supplementary Materials for : Phonon number sensitive electromechanics}}
\\
\end{center}

\end{widetext}

\tableofcontents

\section{Device and fabrication}

\begin{figure} [ht] \centering
\includegraphics[width=1.0\columnwidth]{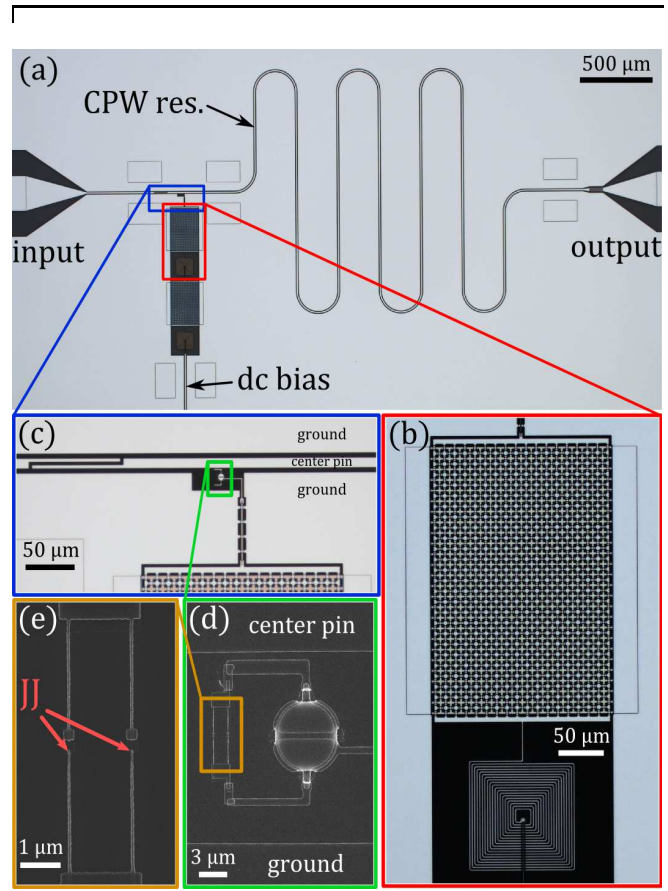}
\caption{ (a) Optical micrograph of the device chip, showing the co-planar waveguide resonator (CPW res.), input and output ports and dc bias input port. (b) One stage of the on-chip LC filter, composed of an inductance in series and a waffled capacitance to ground. (c) Zoom on the area where the qubit and drum are embedded. The input coupling capacitance of the resonator is visible on the left. (d) Scanning electron micrograph of the drum-head and the qubit, also showing the center pin and ground of the CPW resonator. (e) Zoom on the Josephson junctions (JJ). }
\label{SFig_Device}
\end{figure}

Photographs of the full device are given in Fig. \ref{SFig_Device}. Fig. \ref{SFig_Device}(a) shows the overall chip with input and output microwave ports used to probe the co-planar waveguide (CPW) resonator in transmission. This resonator is used to control and readout the qubit. In the following, the microwave resonator is also refered to as the cavity, to avoid possible confusion with the mechanical oscillator. The dc bias port is used to apply dc voltage on the drum, as well as applying rf frequency drives at mechanical frequency and up to 260 MHz. Fig. \ref{SFig_Device}(b) shows the on-chip filter (superconducting LC circuit) that is part of the dc bias line. This is necessary because the dc bias line strongly couples to the qubit mode. Without this filter, the qubit would very rapidly decay into that port (see section \ref{sec:filtering}). As in many standard circuit Quantum Electrodynamics (cQED) setups (e.g. Ref. \cite{Wallraff2004,Riste2015}), we couple the qubit to the CPW resonator by inserting it between the center conductor and one ground plane, as shown in Fig. \ref{SFig_Device}(c). Fig. \ref{SFig_Device}(d) and (e) show scanning electron micrograph of the device, similar to the main text figures.

The fabrication of the device is performed with several steps of aluminium deposition on sapphire. It involves a combination of three optical UV lithography steps performed in a stepper, similar to references \cite{Teufel2011a,Lecocq2015}, as well as two electron beam lithography steps. The optical lithography steps allow the definition of the microwave CPW resonator, the on-chip filter, the drum-head and the qubit islands. The first electron beam lithography step defines the Josephson junctions with the Dolan bridge technique \cite{Dolan1977} (25 nm, then 75 nm thick aluminium at $\pm$20 degree angle). In a second step, we contact the junctions to the rest of the circuit with a bandage technique \cite{Dunsworth2017}. The very last step is the release of the drumhead and all suspended structures (waffle capacitor of Fig. \ref{SFig_Device}(b), air bridges) with a SF6 plasma etch \cite{Teufel2011a,Lecocq2015}. The junctions do not see their critical current ($\approx 10.5$ nA) change significantly during this step.

\begin{table*} [t]
\centering
\begin{tabular}{ l @{\qquad} l }
\toprule
\textrm{Parameter} & 
\textrm{Value} \\
\colrule

qubit maximum Josephson energy & $E_{J,max} \approx 2\pi\times 5.0$ GHz \\
qubit charging energy & $E_C \approx 2\pi\times 2.9$ GHz \\
qubit energy decay time & $T_1 \approx 0.2 - 1$ $\mu$s \\
qubit intrinsic decoherence time & $T_2^* \approx 70 - 95$ ns \\

mechanical frequency & $\omega_m \approx 2\pi\times 25$ MHz \\
mechanical environment temperature & $n_{th} \approx 15$ ($\approx$ 18 mK) \\
mechanical damping rate & $\gamma_m \approx 2\pi\times 94$ Hz \\
mechanical thermal decoherence rate & $n_{th} \gamma_m \approx 2\pi\times 1.4$ kHz \\

microwave resonator (cavity) frequency & $\omega_c = 2\pi\times 4.7574$ GHz \\
microwave resonator (cavity) linewidth & $\kappa \approx 2\pi\times 2.3$ MHz \\

single phonon Stark shift & $2 \chi_m \approx 2\pi\times 0.52$ MHz ($V_d = 6$ V, $n_g=1/2$) \\
qubit - mechanics coupling rate & $g_m \approx 2\pi\times 22$ MHz ($V_d = 6$ V, $n_g=1/2$) \\
qubit - cavity coupling rate & $g_c \approx 2\pi\times 37$ MHz ($n_g=1/2$)\\

\botrule
\end{tabular}
\caption{Summary of the device characteristics}
\label{Table:Param}
\end{table*}

\section{Measurement techniques and basic characterizations}
\begin{figure}[ht] \centering
\includegraphics[width=0.75\columnwidth]{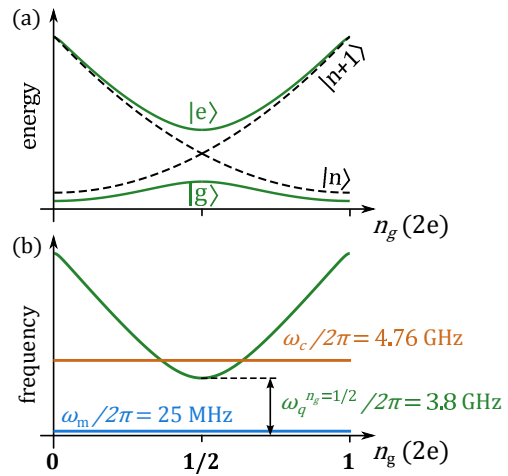}
\caption{ (a) Schematics of the charge qubit energy levels in the Cooper pair box limit. The pure charge states $\ket{n}$ and $\ket{n+1}$ have energies given by the electrostatic charging energy, showing parabolas as a function of gate charge $n_g$. At the charge degeneracy, they hybridize with the Josephson coupling energy to form symmetric and anti-symmetric superpositions of charge states. (b) Frequency landscape of our device as a function of gate charge with $E_J \approx 3.8$ GHz in the absence of qubit-cavity and qubit-oscillator coupling. }
\label{SFig_CPB}
\end{figure}

Before describing how we readout the qubit, we quickly review the working principle of a charge qubit in the Cooper pair box limit \cite{Bouchiat1998,Makhlin2001}, ancestor of the transmon qubit \cite{Koch2007}. The Hamiltonian of the qubit is given by \cite{CottetThesis} 
\begin{equation}
H = 4E_C  \left( \hat{n} - n_g \right)^2 - E_J \cos(\hat{\phi})
\end{equation}
where $E_C$ and $E_J$ are the charging and Josephson energy, $\hat{n}$ is the Cooper pair number operator and $\hat{\phi}$ is the superconducting phase operator. The gate charge parameter $n_g$ is controlled with a local gate voltage. In the Cooper pair box limit $4 E_C \gg E_J$, or at least $E_C \approx E_J$, which is our situation. For most of the data presented, $E_J$ is tuned to $2\pi\times$3.8 GHz with a small external flux, and $E_C \approx 2\pi\times$ 2.9 GHz (see table \ref{Table:Param}). In this limit, the qubit energy levels have a strong dependence on the gate charge $n_g$. As depicted in Fig. \ref{SFig_CPB}, the eigenstates are almost charge states except when $n_g \approx$ 1/2, the point at which ground and excited states are symmetric and anti-symmetric superpositions of charge states. This is the point where the qubit is the most insensitive to charge noise, and where we operate. The qubit frequency is then approximately given by $E_J$, which we can tune using an external flux bias.

\begin{figure}[ht] \centering
\includegraphics[width=0.7\columnwidth]{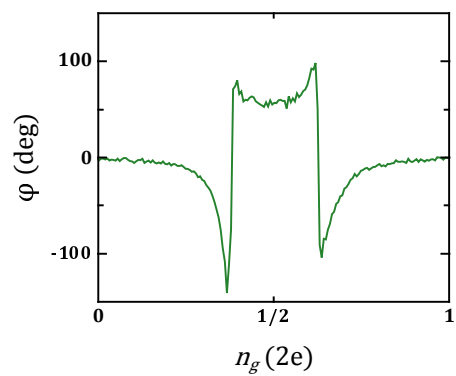}
\caption{ Measurement of the microwave cavity phase shift as a function of gate charge $n_g$ with the qubit in its ground state. This allows the determination of $n_g = 1/2$. When $n_g$ is fixed, the measurement of this phase enables qubit readout. }
\label{SFig_CavityPhase}
\end{figure}

The microwave cavity (i.e. the CPW resonator) is coupled to the qubit such that we operate in the  dispersive limit of the Jaynes-Cummings Hamiltonian \cite{Blais2004} where the cavity frequency is pulled by the state of the qubit. Fig. \ref{SFig_CavityPhase} shows a measurement of the cavity transmission phase measured at the bare cavity frequency, as a function of gate charge. 
This measurement is performed with the qubit in its ground state $\langle \sigma_z \rangle \approx - 1$. It is a direct proxy of the qubit cavity susceptibility $\chi_c$: 
\begin{equation} \label{eq:PhaseShift}
\varphi \approx \frac{2}{\kappa} \operatorname{Re}\left( \chi_c \right) \langle \sigma_z \rangle
\end{equation}
where
\begin{equation} \label{eq:chi_c}
\chi_c = \frac{g_c^2}{-i\Gamma_2^*+\omega_q - \omega_c}
\end{equation}
and $\kappa$ is the (largely over-coupled) cavity linewidth, $g_c$ is the qubit-cavity coupling rate, $\Gamma_2^* = 2/T_2^*$ is the qubit FWHM and $\omega_c$ is the microwave cavity frequency (see table \ref{Table:Param}). Note that $g_c$, $\Gamma_2^*$ and $\omega_q$ depend on the gate charge $n_g$. When the qubit is in its excited state, the phase response given in Fig. \ref{SFig_CavityPhase} has opposite sign. In practice, most of the measurements are done at $n_g \approx 1/2$ and we use the $\langle \sigma_z \rangle$ dependence of $\varphi$ to readout the qubit population. Because our qubit relaxation time $T_1$ is relatively short, we do not have single shot readout of the qubit state. Instead we map out the average value of $\varphi$ on $\langle \sigma_z \rangle = 2 P_e - 1$ \cite{Schuster2005}.

Because there is typically $1/f$ charge noise on the gate parameter $n_g$, we reset it automatically  every one or two minutes to ensure $n_g = 1/2$. This is done by measuring the cavity phase shift given in Fig. \ref{SFig_CavityPhase} and finding the point of symmetry. This reset procedure takes about 6 s.

\begin{figure}[ht] \centering
\includegraphics[width=0.7\columnwidth]{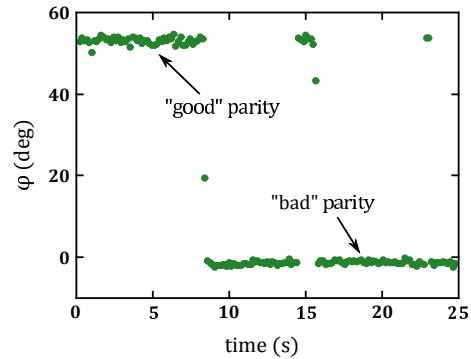}
\caption{ Parity jumps obtained by measuring the cavity phase at $n_g = 1/2$. We run the experiment and measure the qubit conditioned on the parity being in the ``good'' position. }
\label{SFig_Parity}
\end{figure}

Finally, charge qubits are very sensitive to charge number parity. It is known that anomalously large numbers of quasiparticles are present in most superconducting circuits \cite{Aumentado2004,Gunnarsson2004,Shaw2008,Martinis2009,Vool2014}. The tunneling of a quasiparticle through the Josephson junctions causes a sudden change of $n_g$ by 1/2 (that is one electron charge). For a Cooper pair box, such a parity jump takes the qubit at the furthest point away from the sweet spot, dramatically changing the system's properties. We can observe this phenomena by setting $n_g=1/2$ and monitoring the cavity phase in time. Fig. \ref{SFig_Parity} shows such a measurement. This data also demonstrates very good single shot readout of the parity and we estimate our average parity switching time to be about 1 - 10 s. There is no preferred parity in our device. We believe the geometry around our Josephson junction, seen in Fig. \ref{SFig_Device}(e) helps make the parity life time longer. We speculate the long and skinny wires on both sides of the junctions have a much smaller number of quasiparticles than the large superconducting pads of the rest of the device, and crucially make the diffusion of a quasiparticle from one island to the other less likely. In practice, our measurement pulse sequence always contains a readout pulse to readout parity, and we post-select the data in real time based on parity readout with a threshold.

For the spectroscopy data after a sideband drive, presented in Fig. 3 and 4 of the main text, the pulse sequence is the following. We start by turning on simultaneously the sideband drive and the dither (see section \ref{sec:Dither}) and leaving them on for the desired duration. We then wait for 5 $\mu$s to let the qubit relax to its ground state, drive the qubit with a long incoherent pulse (typically 115 $\mu$s) and simultaneously readout the qubit (with less than one average photon in the cavity). We then wait 5 $\mu$s to let the qubit relax again, and measure parity with a long pulse on the cavity (typically 100 $\mu$s, less than one average photon in the cavity). We typically wait 6 ms between each sideband drive to let the mechanical oscillator thermalize back to its environment (see section \ref{sec:gamma_m}).

\section{Time-domain characterization of the bare qubit}
\begin{figure}[ht] \centering
\includegraphics[width=1.0\columnwidth]{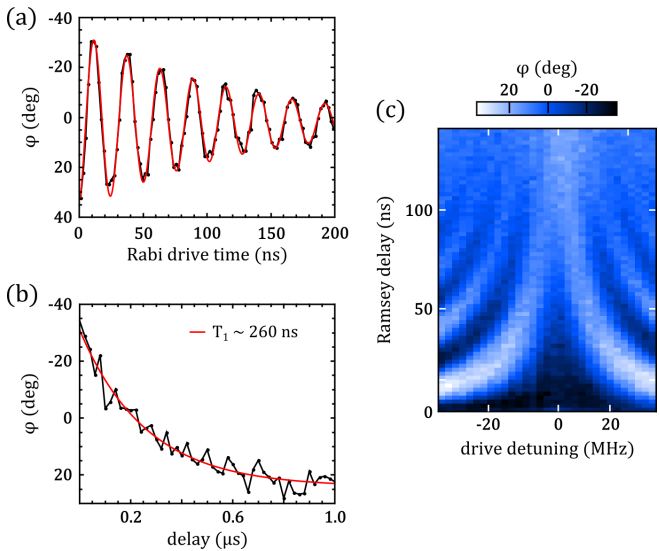}
\caption{ Qubit coherent control. (a) Rabi drive of the qubit. Dots are data and the red line is fit to an exponentially decaying sine. (b) $T_1$ decay of the qubit as a function of delay after a short $\pi$ pulse. Dots are data and red line fits to an exponential decay with 260 ns lifetime. (c) Ramsey fringes as a function of qubit drive detuning. The ramsey characteristic decay time is about $T_2^* \approx 85$ ns, consistent with the low power spectroscopic linewidth of about 3.7 MHz. }
\label{SFig_CohCont}
\end{figure}

Fig. \ref{SFig_CohCont}(a) shows coherent Rabi drive of our qubit transition at $V_d = 0$ V. Fig. \ref{SFig_CohCont}(b) and (c) show $T_1$ and $T_2*$ measurements. $T_2^*$ is relatively stable in time, and can be seen to drop when the coupling is turned on, consistent with the linewidth given in the main text. $T_1$ however depends significantly on the pulse sequence. A long saturating pulse can make our $T_1$ go up to about 1 $\mu$s. We attribute this effect to two-level systems in the residues of silicon nitride between the two plates of drum-head. This is consistent with what was obtained in previous electromechanics experiments with linear LC circuits,  with internal loss going from 150 kHz at high power to 1 MHz at low power \cite{Lecocq2015}.

\section{Measurement of $\chi_m$} \label{sec:chiMeasurement}
The single phonon ac Stark shift $2 \chi_m$ is equal to twice the dispersive shift of the mechanical oscillator due to the qubit in its ground state \cite{Blais2004,Wallraff2004}. Both these effects are widely used in cQED. We can therefore obtain $\chi_m$ by measuring how the mechanical oscillator's resonance frequency is pulled by the interaction with the qubit when the gate charge $n_g$ is being changed. Such a measurement was first realized with a Cooper pair box qubit coupled a silicon nitride beam \cite{LaHaye2009}, and reached $\chi_m \approx 2\pi\times$1.6 kHz. In our setup, we do not have the possibility to independently measure the mechanical oscillator, as it only couples to the qubit (and to a lossy thermal environment). However, we can coherently drive motion on the drum-head and use the qubit, or the qubit-cavity system, to detect the mechanical resonance. \\

\begin{figure}[ht] \centering
\includegraphics[width=1.0\columnwidth]{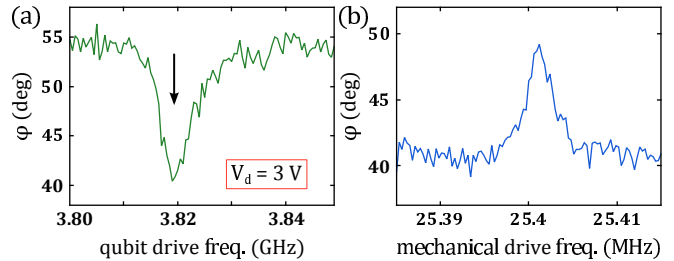}
\caption{ Mechanical oscillator resonance measured at $V_d = 3$ V. (a) Spectroscopy of the qubit at $n_g =1/2$ without mechanical coherent drive. (b) Phase measurement with a qubit spectroscopic tone parked on the qubit resonance (vertical arrow in (a)) as a function of the frequency of the coherent drive on the mechanics. When the mechanical drive hits the mechanical resonance, the qubit is ac Stark shifted and the qubit drive becomes off-resonant. The cavity phase goes back up to its value corresponding to the qubit being in the ground state, hence the observed contrast. The width of the mechanical resonance is broadened by the short pulse used to drive coherent motion. }
\label{SFig_MechSpec}
\end{figure}

%\begin{figure*}[ht] \centering
%\includegraphics[width=1.0\textwidth]{../Figures/SFig_ChiM/SFig_ChiM.ps}
%\caption{ Measuring $\chi_m$. (a) Principle of the measurement. The qubit frequency goes down to its minimum around $n_g=1/2$ (and the transverse coupling goes up). The qubit in its ground state dispersively shifts the coupled cavity and coupled mechanical oscillator, as depicted in the red square inset. Given a mechanical drive at fixed frequency, this drive can be on resonance with the mechanics at specific gate chargesn as indicated by the vertical arrows. (b) Measured cavity transmission phase as a function of gate charge (averaged over both parities). Green dashed line is for no drive on the mechanics (thermal state) and full blue line is when a strong drive at 25.46 MHz has been applied on the mechanics before measuring the phase. As indicated by the vertical arrows, there are resonances appearing symmetrically around $n_g=1/2$. (c) Difference between the signals in (b), making the resonances appear clearly. (d) and (e) Same as in (c), but measured for several drive frequencies. The mechanical resonance is seen to disperse and pulled down around $n_g=1/2$, by an amount close to $\chi_m$. (d) is measured at $V_d = 2$ V and (e) and measured at $V_d = 4$ V. The multiple copies of the mechanical resonance come from the fourier transform of our pulsed drive (our overall pulse sequence is shorter than the mechanical lifetime). }
%\label{SFig_ChiM}
%\end{figure*}

\begin{figure*}[ht] \centering
\includegraphics[width=0.7\textwidth]{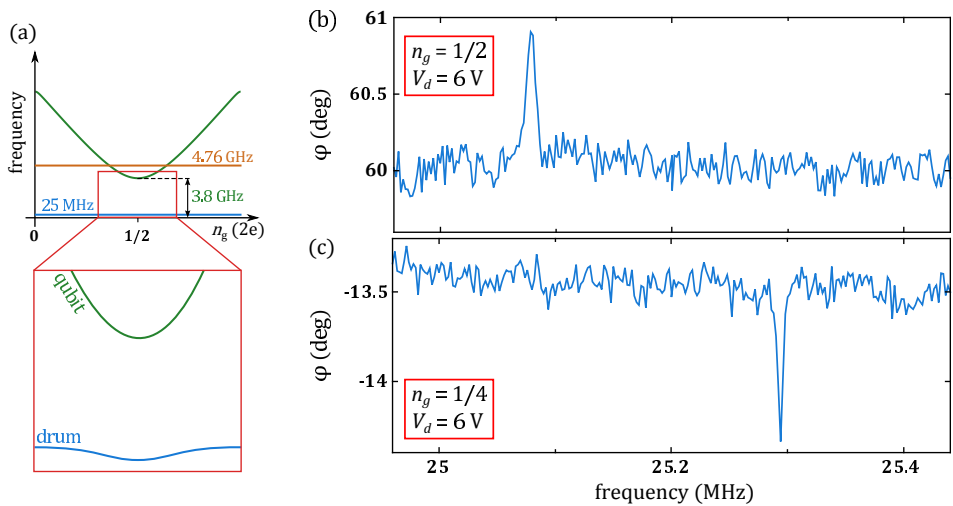}
\caption{ Measuring $\chi_m$. (a) Principle of the measurement. The qubit frequency goes down to its minimum around $n_g=1/2$ (and the transverse coupling goes up). The qubit in its ground state dispersively shifts the coupled mechanical oscillator, as depicted in the red square inset. (b) Measured cavity phase transmission after a 10 $\mu$s drive on the drum at varying frequency with $n_g = 1/2$ (and $V_d = 6$V). When the drive frequency hits the resonance of the mechanics, motion is driven on the drum, which in turns rectifies the phase signal and allows spectroscopy of the oscillator. (c) Same as (b), but measured at $n_g = 1/4$, showing a significantly less shifted mechanical frequency. }
\label{SFig_ChiM}
\end{figure*}

Motion is driven resonantly with the product of the dc voltage $V_d$ and an ac voltage drive applied on the same port (see Fig. \ref{SFig_Wiring}). Once the coherent drive is turned off, motion survives for a characteristic time scale given by the mechanical damping rate, approximately $1.7$ ms (see section \ref{sec:gamma_m}). During that time, the qubit gate charge $n_g$ is being modulated at $\omega_m$ with an amplitude given by mechanical motion. This modulation is effectively an ac Stark shift on the qubit, as explained in details in the main text, but it also rectifies the cavity phase signal. This can be understood by looking at equation \ref{eq:PhaseShift} or by considering the phase signal of Fig. \ref{SFig_CavityPhase} with an ac modulation on $n_g$ due to mechanical motion. Note that the qubit linewidth is also dressed by this effect, and this enters equation \ref{eq:PhaseShift} via the qubit-cavity susceptibility $\chi_c$ (equation \ref{eq:chi_c}). \\

In the low motion amplitude limit, the rectification is small, and we detect mechanical resonance via qubit spectroscopy. A typical measurement of this kind is shown in Fig. \ref{SFig_MechSpec}. This works well at $n_g = 1/2$ because this is where the qubit spectroscopy is the narrowest. However charge noise makes the qubit resonance extremely broad away from $n_g=1/2$ \cite{Blais2004}. In order to measure the mechanical resonance away from $n_g = 1/2$, we drive a sufficiently large motion amplitude on the drum so that the cavity phase signal is modified even though the qubit stays in its ground state. The amplitude of this rectification depends on the curvature of the phase signal. Fig. \ref{SFig_ChiM} shows such measurements. At $n_g = 1/2$, we observe a resonance at around 25.078 MHz in the phase response of the cavity (selecting on charge parity). At $n_g = 1/4$, we observe a resonance at around 25.294 MHz (both parities are degenerate and we do not select on parity). In those measurements, the resonance linewidth is broadened by the length of the drive pulse (10 $\mu$s). The background phase value and the sign of the contrast on resonance can be understood by looking at the value and the curvature of the phase signal given in Fig. \ref{SFig_CavityPhase}. At $n_g = 0$ (or 1), the mechanical frequency should be very close to the bare mechanical frequency. However the very tiny curvature of the phase signal at $n_g = 0$ holds us from taking data there with reasonable phonon number in the mechanics. Instead, we account for the fact that the dispersive shift is non zero at $n_g=1/4$ to calculate $\chi_m$ at $n_g = 1/2$. We have checked the power dependence of our $\chi_m$ measurement and the quoted value is that obtained in the low phonon number limit where there is no power dependence (about 200 phonon from the coherent drive). In comparison, we estimate the critical number of phonons $\omega_q/4\chi_m^2$ \cite{Blais2004} to be about $1.4\times 10^4$ at $n_g = 1/2$. The obtained dispersive shift $\chi_m \approx 2\pi\times 0.26$ MHz is a 160 fold improvement compared to Ref. \cite{LaHaye2009}.

\section{Expected $\chi_m$ from simulations, expected maximum bias voltage $V_d$. }
%Figure \ref{SFig_ChiM_Vd} shows the dependence of the experimental single phonon Stark shift $2\chi_m$ (black dots), obtained with the technique described in section \ref{sec:chiMeasurement}, as a function of the bias voltage $V_d$. We can compute the expected $2\chi_m$ from a circuit model that includes the static displacement of the drum-head as well as parasitic capacitances obtained from finite element simulations. This model yield the solid lines in figure \ref{SFig_ChiM_Vd}. For the purpose of clarity, we show here the derivation of $2\chi_m$ for a given static displacement, and without the parasitic capacitances. We start with the circuit given figure figure 1 in the main text. This circuit makes the wrong assumption that both capacitances to ground are the same, $C_0$, but this is equivalent to neglecting parasitics. We write the drum capacitances as 

We can compute the expected $2\chi_m$ from a circuit model that includes the static displacement of the drum-head as well as parasitic capacitances obtained from finite element simulations. For the purpose of clarity, we start here with the derivation of $2\chi_m$ for a given static displacement, and without the parasitic capacitances. We start with the circuit given Fig. 1 of the main text. This circuit makes the simplifying assumption that both capacitances to ground are the same, $C_0$, which is also equivalent to neglecting parasitics. We write the drum capacitances as 
\begin{equation}
C_{\textrm{drum}}^\pm = \frac{C_d}{1 \mp \frac{x}{l}} \approx C_d (1 \pm \frac{x}{l})
\end{equation}
where $l$ is the static separation of the drum plates (the vacuum gap). The Hamiltonian of the system writes
\begin{equation}
H = 4 E_C \left( \hat{n} - n_g \right)^2 + E_J \cos \hat{\phi} + \frac{1}{2} m \omega_m \hat{x}^2 + \frac{\hat{p}^2}{2 m}
\end{equation}
where $m$ is the mass of the oscillator, $\hat{x}$ is the position operator and  $\hat{p}$ is the momentum operator. $E_c$ and $n_g$ are here functions of the position of the oscillator and this is how the qubit couples to the oscillator. Because the expected gap $l\approx 50$ nm, and the zero point motion of the oscillator $x_{zpf}\approx 3.6$ fm, expanding to first order in $x/l$ is very good approximation even at large numbers of quanta. We therefore write
\begin{equation}
H \approx H_0 + \frac{\partial H}{\partial x} \hat{x}
\end{equation}
where $H_0$ is the static (decoupled) Hamiltonian. At the charge degeneracy point : 
\begin{equation}
\frac{\partial H}{\partial x} \hat{x} = e \beta_m V_d \frac{1}{l} \hat{n} \hat{x} = g_m \sigma_x (a + a^\dag)
\end{equation}
Where $e$ is the electron charge, $a$ is the phonon annihilation operator, and
\begin{align}
g_m & = e \beta_m V_d \frac{x_{zpf}}{l} \\
\beta_m & = \frac{4C_0 C_d}{(C_0 + C_d)(C_0+C_d+2 C_J)}
\end{align}
From $g_m$ we calculate the single phonon Stark shift, including the Bloch-Siegert shift \cite{Bloch1940,Zueco2009,Beaudoin2011}, and in the limit $\omega_m \ll \omega_q$ :
\begin{equation}
2 \chi_m \approx \frac{4 g_m^2}{\omega_q}
\end{equation}

The static position $l$ at any fixed $V_d$ was estimated semi-analytically by finding the minimum of the potential energy $U$ given by the sum of the elastic restoring force of the oscillator and the electrostatic force :
\begin{equation} \label{eq:Potential}
U(l) = \frac{1}{2} k (l-l_0)^2 - \frac{1}{2}\frac{1}{\frac{1}{C_0}+\frac{l}{\epsilon_0 S}} V_d^2
\end{equation}
where $k\approx$ 190 N$\cdot$m is the effective spring constant, $l_0$ is the gap at zero voltage, $\epsilon_0$ is the vacuum permittivity and $S$ is the area of the drum. Finally, the parasitic capacitances are estimated with \emph{Sonnet} and included in a modifed version of the above model. Our measured value $2 \chi_m = 2\pi\times$0.52 MHz at $V_d =$ 6 V indicates a gap of about 56 nm. This is consistent with our expectations based on the gap estimated in similar devices \cite{Teufel2011}, of about 50 nm.
%As shown in figure \ref{SFig_ChiM_Vd}, the agreement between data and simulation is rather good, but the comparison seems to suggest a gap at $V_d =$0 V around 70 or 80 nm. This a little bit far from the design value of 50 nm, but not impossible. The measured $\chi_m$ could also be underestimated in our technique requiring large phonon occupation. The simulation could also be underestimating the parasitic capacitances, which would reduce the interaction strength.

%\begin{figure}[ht] \centering
%\includegraphics[width=0.8\columnwidth]{../Figures/SFig_ChiM_Vd/SFig_ChiM_Vd.ps}
%\caption{ $\chi_m$ dependence on the bias voltage $V_d$. Dots are data and solid lines are theory without fitting parameters, but for several different gaps between bottom and top electrodes of the drum-head (gap at $V_d =$0 V). The theory lines were produces using numerical simulation to obtain all the parasitic capacitances and include the static displacement of the drum with $V_d$ as well as the voltage instability. The final points on the simulation curves correspond to the expected collapse voltages. The horizontal dashed line is the linewidth of the bare qubit in the present device. }
%\label{SFig_ChiM_Vd}
%\end{figure}

The device was designed to be able to support high dc voltages, limited by the instability present in the potential of equation \ref{eq:Potential}, expected at 21V. We have tested devices for which the electrodes underneath the drum-head were effectively grounded at dc (this could be the case with a qubit if we used a single island Cooper pair box, with the Josephson junction being a short circuit to ground at dc). The maximum voltage we could apply on those devices is about 3.2 V and matches our model of equation \ref{eq:Potential}. This is the voltage at which the electrostatic instability makes the drum collapse. Here we designed small capacitances in series to ground ($C_0$, see main text), which reduces the coupling $g_m$ per volt, but favorably increases the maximum voltage to 20 - 25 V (depending on the gap $l_0$). This transforms the voltage source $V_d$ into a charge source, with the voltage instability disappearing when $C_0 < C_{\textrm{drum}}/2$. Note that in reference \cite{Andrews2015} this collapse voltage is close to 18 V, but only because the dc voltage bias is applied on a small area of the drum, making the effective capacitance much smaller. In practice, the highest voltage we can apply is limited by our ability to readout the qubit. Around $V_d = 8$ V, the phase signal of the cavity becomes bi-stable (within the ``good'' parity position) even though the qubit is not driven externally. Understanding and taming this effect would allow us to push $2\chi_m$ up, expectedly to $2\pi\times$10 MHz at $V_d \approx$ 21 V. Such Stark shift would possible exceed the linewidth of the qubit and place the device in the strong dispersive limit \cite{Schuster2007a}.

\section{$P(n)$ reconstruction technique and bootstrapping}
\subsection{Dressed qubit spectrum approximation (equation 2 of the main text)}
As mentioned in the main text, the validity of equation 2 of the main text, which we recall here:
\begin{equation} \label{eq:Convolution}
P_e^{\textrm{dressed}}(\omega) = \sum_n P(n) P_e^{\textrm{bare}}(\omega-2\chi_m n)
\end{equation}
depends on three conditions. First, the system must be in the weak damping limit $\gamma_m \ll \chi_m$, with $\gamma_m$ the mechanical damping rate. This condition ensures weak back-action \cite{Clerk2007a}, so that the height of each underlying peak is $P(n)$. Second, the width of each underlying peak $n$ (approximately given by $\Gamma_2^* + 2 n n_{th} \gamma_m$ \cite{Clerk2007a}, $n_{th}$ being the average thermal occupancy) is dominated by the bare qubit linewidth $\Gamma_2^*$. Third, the number of phonons is small compared to $\omega_m / 2 \chi_m$ to avoid unwanted degeneracy between dressed qubit transitions and second order sideband transitions. More precisely, a pure qubit transition at $\omega_q + 2 \chi_m n_A$ can be at the same frequency as a sideband transition at $\omega_q + \omega_m + 2 \chi_m (n_B + 1/2)$ when $n_A = n_B + \omega_m/2 \chi_m$. These sidebands are not the ones we use to control the phonon number distribution (see section \ref{sec:Dither}). When driven with the spectroscopic tone, they have a small contribution to the qubit spectrum, as seen in Fig. 2 and 3 of the main text. 

We experimentally determine $\gamma_m \approx 2 \pi \times$ 94 Hz (section \ref{sec:gamma_m}) and $n_{th} \approx 15$ (see main text). This ensures the first condition to be very well satisfied. The second condition is valid until about 300 phonons and corrections can easily be added to equation 2 to account for broadening (see section \ref{sec:Fig2&3SM}). The third condition is only valid up to about 50 phonons, but modifying equation 2 to account for large phonon numbers is a more challenging task which is beyond the scope of this paper. This holds us from being quantitative at large phonon numbers but allows us to make qualitative statements about arbitrary distributions, as shown in the main text.

% We can phenomenologically understand that the picture given in figure 2(a) of the main text breaks down when $2 \chi_m n \approx \omega_m$ in the following way. Consider the response of the Cooper pair box to a classical field driving its gate charge at $\omega_{\textrm{mod}}$. Such a drive performs non-linear frequency modulation of the qubit resonance. As for any frequency modulation, significant sidebands begin to appear on the signal when the carrier frequency is modulated by an amount comparable to $\omega_{\textrm{mod}}$. Because mechanical motion at $V_d \neq 0$ is essentially equivalent to a gate charge modulation of the qubit at $\omega_m$, this means strong sidebands should begin to appear on the qubit spectrum for Fock numbers such that $2 \chi_m n \approx \omega_m$. Because our oscillator is low frequency, this happens relatively quickly for our device, at $n \approx 50$ phonons. 

\subsection{Bayesian-based estimation of $P(n)$} \label{sec:LR}

\begin{figure}[ht] \centering
\includegraphics[width=1.0\columnwidth]{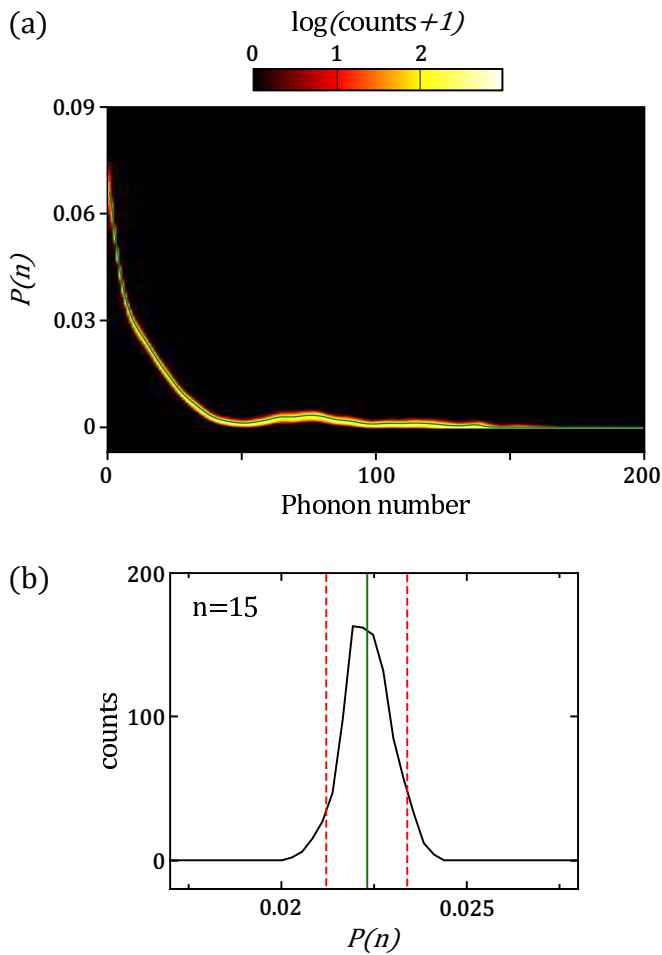}
\caption{ Reconstruction bootstrap and confidence intervals. (a) Histograms of the bootstrap distributions for the thermal state shown in the main text represented in color scale (log scale). The green solid line is the estimated $P(n)$. (b) An example of histogram at $n=15$. The black curve is a histogram of the bootsrap distribution (linear scale), the vertical green line is the estimated $P(n=15)$ and the dashed red lines are the 5\% confidence intervals. }
\label{SFig_Bootstrap}
\end{figure}

We now describe the algorithm that we use to extract the mechanical oscillator energy distribution $P(n)$ from the measured dressed qubit spectroscopy. This is essentially an adapted version of the basic Lucy-Richardson algorithm often used in astronomy \cite{Richardson1972,Lucy1974}. The Lucy-Richardson algorithm is an iterative, Bayesian-based method, which was originally aimed at reconstructing images degraded by optical aberrations. It is essentially a deconvolution procedure. It is relatively protected against noise and we verify this with a non-parametric bootstrap. The important difference between the original Lucy-Richardson and our algorithm is the fact that we have important extra information on the convolution function : we know it is the bare qubit shifted by $2 \chi_m n$ and we are only looking for the height of each underlying peak. 

More precisely, we can re-write our measured dressed qubit spectroscopy as in the main text, but in more practical notations :
\begin{equation} \label{eq:Convolution}
S_i = \sum_n P(n) \Pi_{ni}
\end{equation}
where $S_i$ is the measured signal at the frequency bin $i$ (excited state probability of the qubit at that particular drive frequency), $P(n)$ is the phonon population on the number state $n$ and $\Pi_{ni}$ is the equivalent of a point spread function, or convolution function. Importantly, $\Pi_{ni}$ contains here much more information than in the case where this is a convolution by imperfect optics. In particular it contains information about the positions of each number state $n$ in the frequency space as well as the bare qubit linewidth and the Rabi drive frequency. For a given number state $n$, $\Pi_{n i}$ is a Lorentzian (or skewed Lorentzian, see section \ref{sec:Fig4SM}) centered around $\omega_q + 2 \chi_m n$, with FWHM and height determined by the qubit characteristics and the drive strength. To invert equation \ref{eq:Convolution} and obtain $P(n)$, we use the Lucy-Richardson recurrence formula
\begin{equation} \label{eq:Deconvolution}
P_{k+1}(n) = P_k(n) \sum_i \Pi_{ni} \frac{S_i}{\sum_q P_k(q) \Pi_{qi}}
\end{equation}
where $k$ is the iteration index. This requires a starting guess $P(n)$, which we chose to be a uniform $P(n)=1/N$, where $N$ is the number at which we truncate the Fock space. This choice is equivalent to using Bayes's postulate \cite{Richardson1972}. We checked that a different choice of $P(n)$ might change how many iterations are necessary but it has negligible influence on the result of the reconstruction. In the reconstructions presented in the main text, we run the algorithm for 15 iterations in Fig. 2 and 3 of the main text, and 30 iterations in Fig.4 of the main text. In general, more iterations means a distribution that catches finer details of the data at the expense of adding noise in the $P(n)$ reconstruction. This noise is characterised with a bootstrap (see below) and is shown with the lighter color shades in all presented reconstructions. At each iteration, we ensure $P(n) \geq 0$ and $\sum_n P(n) = 1$ by renormalizing $P(n) \rightarrow P(n)/\sum_n P(n)$. In the data presented, we truncate the Fock space at $N=200$. \\

\subsection{Bootstrap}
To investigate how noise in the data $S_i$ is transformed by our algorithm into noise on $P(n)$, we perform a non-parametric bootstrap \cite{Efron1994}. The data presented in the main text is the result of the average of 200 traces. To create a bootstrap sample, we re-sample randomly with replacements among those 200 traces, average and run the reconstruction. We repeat this procedure for 1000 samples to obtain bootstrap distributions. An example of such distributions is given in Fig. \ref{SFig_Bootstrap}. Based on these distributions, we compute the so-called basic bootstrap 5\% confidence intervals \cite{Efron1994}, as shown in Fig. \ref{SFig_Bootstrap}(b). They are reported in the main text as light shades around the $P(n)$ estimates. Similar to the large signal to noise ratio of the data, the confidence intervals obtained by the bootstrap are close to the mean value. So the noise in the data is well tolerated by the reconstruction algorithm. Instead of data noise, or noise added by the reconstruction, the error in our estimated $P(n)$ is dominated by systematic errors arising from two main sources. The first one is the deviation from the low phonon number approximation discussed above, which is more manifest when we drive blue sideband transitions. The second one is the uncertainty on $\Pi_{ni}$, as discussed below, which dominates when driving red sideband transitions. 

\subsection{Thermal state fit and deconvolution parameters for the data presented in Fig. 2 and 3 of the main text} \label{sec:Fig2&3SM}
The fit of the qubit spectrum dressed with the mechanical oscillator in a thermal state is done with 3 free parameters : the average thermal occupancy, the bare qubit frequency at 6 V, and the Rabi rate. We use the extracted Rabi rate and qubit frequency to perform the deconvolution (along with the measured ac-Stark shift $2\chi_m$ and qubit intrinsic linewidth $\Gamma_2^*$). In particular, the Rabi rate is used to account for spectroscopic power broadening of the intrinsic qubit lineshape of about 1.1 MHz for the data shown in Fig. 2 and 3 of the main text.

\subsection{Sources of uncertainty for $P(n)$ after a red sideband drive (Fig. 4 of the main text)} \label{sec:Fig4SM}

\begin{figure}[ht] \centering
\includegraphics[width=1.0\columnwidth]{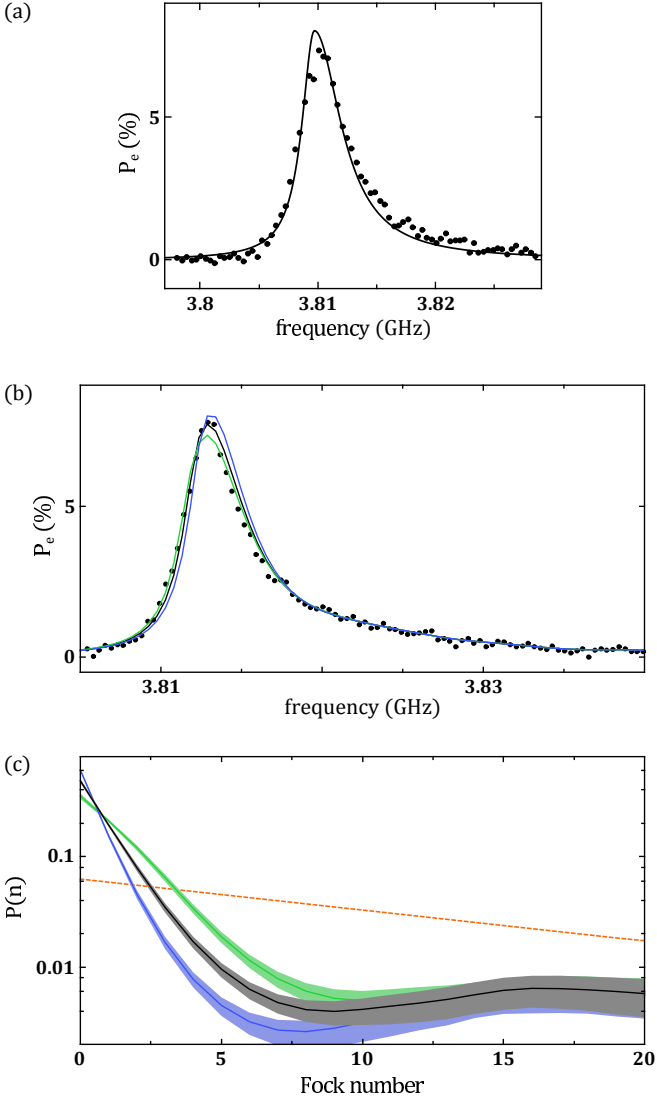}
\caption{ (a) Bare qubit response at $V_d = 0$ V (dots) fitted with a skewed Lorentzian (solid line). (b) Qubit spectroscopy taken after a long red sideband pulse (same as the black data in Fig. 4 of the main text). Solid lines are qubit lineshapes expected from the reconstructed phonon number distribution for different value of estimated bare qubit frequency at $V_d = 6$ V ($f_{q,6V} =$ 3812.01 (green), 3812.37 (black) and 3812.73 (blue) MHz)  (c) Associated phonon distributions. The dashed line shows the thermal state distribution as a reference. }
\label{SFig_PSF-RSB}
\end{figure}

When trying to determine $P(n)$ with a resolution better than $\Gamma_2^*/2\chi_m$, the signal to noise of the data is increasingly important, but what matters even more is the uncertainty on the bare qubit response $P_e^\textrm{bare}(\omega)$ (or equivalently the point spread function $\Pi_{ni}$, see section \ref{sec:LR}). This becomes important for the data presented in Fig. 4 of the main text, for which we estimate the phonon distribution after a red sideband drive. For this set of data, we use the bare qubit response measured at $V_d = 0$ V with the same spectroscopic power and we account for charge noise. Low frequency charge noise is expected to yield an asymmetric lineshape due to the dispersion relation of the qubit at $n_g=1/2$ (see Fig. \ref{SFig_CPB}). As shown in Fig. \ref{SFig_PSF-RSB}(a), we approximate this with a skewed Lorentzian 
\begin{equation}
P_e^\textrm{bare}(\omega)=\frac{1}{2} \frac{ \left( A\Gamma_2^*/2 \right)^2 }{ \left( \frac{\omega-\omega_{q,0V}}{1+L \textrm{sgn}(\omega-\omega_{q,0V})} \right)^2 + \left( \frac{\Gamma_2^*}{2} \right)^2 (1+A^2)  }
\end{equation}
We now fit the bare qubit response ($V_d =0$ V) with 3 free parameters being the reduced Rabi rate $A=\Omega_{R}\sqrt{\frac{2}{\Gamma_1 \Gamma_2^*}}$, the asymmetry $L$ and the qubit frequency $f_{q,0V}=\omega_{q,0V}/2\pi$. When the phonon distribution is mainly localized around $n=0$, the uncertainty on the qubit frequency completely dominates the error on $P(n)$. Going from $V_d = 0$ V to 6 V, we estimate the small static displacement of the drum to be 1.5 nm. This changes the charging energy of the qubit which slightly changes the qubit frequency at $n_g=1/2$ by about - 0.5 MHz. This is comparable to the Lamb (and Bloch-Siegert) shift $\chi_m$ of + 0.26 MHz. Overall, it is difficult to be confident about the bare qubit frequency at $V_d = 6$ V with much less than 1 MHz uncertainty. We therefore estimate $f_{q,6V}$ by measuring several qubit spectroscopies at 6 V with the oscillator in a thermal state and at different spectroscopy powers. We then fit those spectroscopies leaving $f_{q,6V}$ as a free parameter (along with $n_{th}$, fixing all other parameters from the qubit measured at $V_d = 0$ V) and we obtain $f_{q,6V} = 3812.37 \pm 0.36$ MHz. This uncertainty is reponsible for the uncertainty on $P(0) = 0.48\pm 0.13$ in Fig. 4 main text. Fig. \ref{SFig_PSF-RSB}(b) shows reconstructed $P(n)$ with $f_{q,6V} =$ 3812.01, 3812.37 and 3812.73 MHz, for a qubit spectroscopy taken after a 1.5 ms red sideband pulse driven around $n=8$ (see main text). This data explicitly shows our uncertainty on $P(n)$. Finally, note that the entire data set of Fig. 4 was taken at a slightly different flux from the data presented in Fig. 2 and 3.

\section{AC dither sideband transitions} \label{sec:Dither}

\begin{figure}[ht] \centering
\includegraphics[width=1.0\columnwidth]{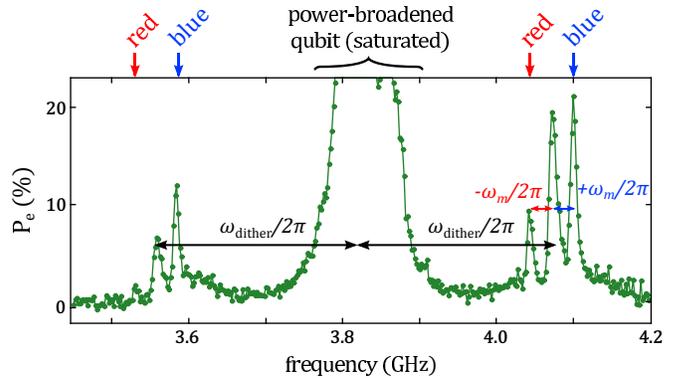}
\caption{ Sideband transition spectrum at large probe power, $V_d = 6$ V. The indicated red and blue sideband transitions are those used in the main text to manipulate the phonon populations. }
\label{SFig_SBSpectrum}
\end{figure}

At the charge degeneracy of the Cooper pair box $n_g=1/2$, the symmetry of the Hamiltonian requires conservation of the number of excitations \cite{Blais2007}. This is at least true for the Jaynes-Cummings interaction $g_m ( \sigma_+ a + \sigma_- a^\dag)$ \cite{Beaudoin2011}. This yields to selection rules on the sideband transitions between the qubit, the mechanical oscillator and the microwave drive \cite{Blais2007,Beaudoin2011}. In particular, 
\begin{itemize}
\item a first order sideband (that takes only one photon at a time from the microwave drive) must be a two-phonon process $\ket{g,n} \leftrightarrow \ket{e,n\pm 2}$ and goes at the rate $\Omega_{SB}^{\pm 2} \approx \chi_m \Omega_R/\Delta_{qd}$.
\item a second order sideband (takes two drive photons) will be a one phonon process  $\ket{g,n} \leftrightarrow \ket{e,n\pm 1}$ at the rate $\Omega_{SB}^{\pm 1} \approx g_m (\Omega_R/\Delta_{qd})^2$
\end{itemize}
where $\Omega_R$ is the Rabi rate of the drive and $\Delta$ is the detuning between the drive and the qubit. These rates are calculated under the approximation that $\Omega_R / \Delta_{qd} \ll 1$. Therefore $\Omega_{SB}^{\pm 2}$ is a slow rate in our case. $\Omega_{SB}^{\pm 1}$ can in principle be faster but driving a second order sideband requires either tones with very large power, or tones very close to the mechanical resonance and the qubit resonance to maintain $\Omega_R / \Delta_{qd} \approx 0.1 $. This is challenging because even though we are in the sideband-resolved regime ($\omega_m > \Gamma_2^*$), drive power quickly makes the qubit transition broader than $2 \omega_m$. Instead, we drive the so-called ac-dither sidebands introduced in reference \cite{Blais2007} for a Cooper pair box. They rely on the fact that the qubit-mechanics coupling is purely transverse at $n_g=1/2$, but has non-zero longitudinal component (along $\sigma_z$) as soon as $n_g\neq 1/2$. Applying a weak dither on the gate charge, $n_g(t) = n_{g0} + n_g^{\textrm{dither}} \cos\left( \omega_{\textrm{dither}} t \right)$, dynamically enables a longitudinal coupling rate of the form $\sigma_z (a + a^\dag)$. This changes the selection rules and allows us to drive a first order process $\ket{g,n} \leftrightarrow \ket{e,n\pm 1}$ at the rate given in equation \ref{eq:SBrate}. Although this makes the qubit spend some time away from the charge degeneracy point, it does not degrade the dephasing time due to charge noise. This is because the added dephasing is echoed away thanks to the periodicity of the drive and the symmetry of the qubit dispersion relation \cite{Blais2007}.

Fig. \ref{SFig_SBSpectrum} shows the full measured spectrum of the sideband transitions, with dither turned on at $\omega_{\textrm{dither}} =2\pi\times$260 MHz and large microwave probe power. The qubit transition is power broadened and saturated, with $P_e \approx 0.5$ on resonance. Two groups of 3 satellite peaks are visible on either side of the qubit, detuned by the dither frequency. In the main text, we omitted $\omega_{\textrm{dither}}$ in the frequency of the sideband transitions for clarity. Keeping this term, as described in reference \cite{Blais2007}, we identify the blue and red sideband transitions at 
\begin{align}
\omega_B & = \omega_q \pm \omega_{\textrm{dither}} + \omega_m + 2\chi_m \left( n + \frac{1}{2} \right) \\
\omega_R & = \omega_q \pm \omega_{\textrm{dither}} - \omega_m + 2\chi_m \left( n - \frac{1}{2} \right)
\end{align}
\\

\section{Modeling : semi-classical master equation, sideband rates and numerical simulation}
\subsection{Principle}
We consider $N$ states in the Fock basis of the mechanical oscillator and the ground and excited state of the qubit. We take those $2N$ independent states and assume that all drives are classical so that the probabilities evolve under a classical master equation, similar to classical laser rate equations. This approximation is valid because we drive all our sidebands on a time scale much longer than the qubit dephasing time ($\approx 80$ ns) and relaxation time ($\approx 260$ ns) and with a rate $\Omega_{SB} \approx 2\pi\times 55$ kHz (see section \ref{sec:SidebandRate}) much smaller than qubit dephasing rate $\Gamma_2^* \approx 2\pi\times 3.7$ MHz. In other words, our density matrix is entirely diagonal. The simulations given in Fig. 3 (blue sideband drive) are based on the numerical time integration of the coupled differential equations (for all $n$ from $0$ to $N$) :
\begin{widetext}
\begin{align}
\frac{d}{dt} P_g(n) & = \gamma_m n_{th} n P_g(n-1) + \gamma_m (n_{th}+1) (n+1) P_g(n+1) - \gamma_m \left( n_{th} (n+1) + (n_{th}+1) n \right) P_g(n) \nonumber  \\
\quad & -\Gamma_{SB}(n,n_{\textrm{drive}}) P_g(n) + \Gamma_{SB}(n,n_{\textrm{drive}}) P_e(n+1) \nonumber \\
\quad & + \Gamma_1 P_e(n) \label{eq:Master1} \\ 
\frac{d}{dt} P_e(n) & = \gamma_m n_{th} n P_e(n-1) + \gamma_m (n_{th}+1) (n+1) P_e(n+1) - \gamma_m \left( n_{th} (n+1) + (n_{th}+1) n \right) P_e(n) \nonumber \\
\quad & + \Gamma_{SB}(n-1,n_{\textrm{drive}}) P_g(n-1) - \Gamma_{SB}(n-1,n_{\textrm{drive}}) P_e(n) \nonumber \\
\quad & -\Gamma_1 P_e(n) \label{eq:Master2}
\end{align}
\end{widetext}
%\begin{align}
%\frac{d}{dt} P_g(n) & = \gamma_m n_{th} n P_g(n-1) \nonumber \\ 
%& \quad + \gamma_m (n_{th}+1) (n+1) P_g(n+1) \nonumber \\
%& \quad - \gamma_m \left( n_{th} (n+1) + (n_{th}+1) n \right) P_g(n)   \\
%\frac{d}{dt} P_g(n) & = -\Gamma_{SB,n} P_g(n) + \Gamma_{SB,n} P_e(n+1) \\
%\frac{d}{dt} P_e(n) & = \Gamma_{SB,n-1} P_g(n-1) - \Gamma_{SB,n-1} P_e(n) \\
%\frac{d}{dt} P_e(n) & = -\Gamma_1 P_e(n) \\
%\frac{d}{dt} P_g(n) & = \Gamma_1 P_e(n) 
%\end{align}
Where $P_{g(e)}(n)$ is the population in the qubit ground (excited) state with $n$ phonons, $\Gamma_1=1/T_1$, and $\Gamma_{SB}$ is the sideband rate. In this set of equations, we can identify the dynamics coming from the thermal bath of the mechanical oscillator \cite{Milburn} (1st line of each equation), the number-sensitive sideband drive (2nd line) and the qubit decay $T_1$ (3rd line). To compute the number dependence of the blue sideband drive, we start with the ac-dither sideband Hamiltonian of Ref. \cite{Blais2007} :
\begin{equation}
H_{BSB}/\hbar = \Omega_{SB,0} \left( a \sigma_- + a^\dag \sigma_+ \right)
\end{equation} 
where $a$ is the phonon annihilation operator, $\sigma_-$ is the qubit lowering operator and $\Omega_{SB,0}$ is given by 
\begin{equation} \label{eq:SBrate}
\Omega_{SB,0} = g_m \frac{\Omega_R}{2(\omega_{\textrm{drive}}-\omega_q)} J_1 \left( \frac{8 E_C n_g^{\textrm{dither}}}{E_J} \right)
\end{equation}
where $\Omega_R$ is the Rabi rate given by the power of the microwave drive, $J_1$ is the 1st Bessel function and $n_g^{\textrm{dither}}$ is the amplitude of the dither. To obtain our sideband rate $\Gamma_{SB}$, we identify the steady state solution of the quantum equations of motion obtained from $H_{BSB}$ with the associated semi-classical rate equations. For all $n$, in the subspace $\left\{ \ket{g,n}, \ket{e,n+1} \right\}$, the steady state solution of the quantum equation of motion is
\begin{equation}
P_{\ket{e,n+1}} = \frac{1}{2 + \frac{\Gamma_1 \Gamma_2^*}{4 (n+1) \Omega_{SB,0}^2} \frac{\Delta^2+(\Gamma_2^*/2)^2}{(\Gamma_2^*/2)^2}}
\end{equation}
where $\Gamma_2^*=2/T_2^*$ and $\Delta=\omega_{B}(n)-\omega_{\textrm{drive}}$ is the detuning between the blue sideband transition frequency and the drive frequency. The analog rate equation solution is
\begin{equation}
P_{\ket{e,n+1}} = \frac{1}{2+\frac{\Gamma_1}{\Gamma_{SB}}}
\end{equation}
We therefore identify the sideband rate to be 
\begin{align}
\Gamma_{SB}(n,n_{\textrm{drive}}) & = \frac{4 (n+1) \Omega_{SB,0}^2}{\Gamma_2^*} \frac{(\Gamma_2^*/2)^2}{\Delta^2+(\Gamma_2^*/2)^2} \\
& = \frac{4 \Omega_{SB,0}^2}{\Gamma_2^*} \frac{n+1}{1+ \left( \frac{4 \chi_m ( n-n_{\textrm{drive}} )}{\Gamma_2^*} \right)^2 }
\end{align}

For drives centered on numbers sufficiently far from 0, $n_{\textrm{drive}} > \Gamma_2^*/2\chi_m \approx 7$, the sideband rate $\Gamma_{SB}(n,n_{\textrm{drive}})$ is well approximated by a Lorentzian centered on $n_{\textrm{drive}}$ with a FWHM given by $\Gamma_2^*/2\chi_m$. We therefore drive on the order of 7 transitions. For a drive at $n_{\textrm{drive}}=0$, the sideband rate is not a Lorentzian; it has a FHWM of about 10.

Note that in the case of a red sideband drive, $\Gamma_{SB}(n,n_{\textrm{drive}})$ writes $n$ at the numerator instead of $n+1$, in principle yielding to sideband asymmetry \cite{Weinstein2014}:
\begin{equation}
\Gamma_{SB,\textrm{red}}(n,n_{\textrm{drive}}) = \frac{4 \Omega_{SB,0}^2}{\Gamma_2^*} \frac{n}{1+ \left( \frac{4 \chi_m ( n-n_{\textrm{drive}} )}{\Gamma_2^*} \right)^2 }
\end{equation}
In the numerical simulation of the set of equations \ref{eq:Master1} and \ref{eq:Master2}, we ensure $\sum_n P_g(n)+P_e(n) = 1$ at each time step of the simulation by setting $P_g(N)+P_e(N)=1-\sum_{n=0}^{N-1} \left( P_g(n)+P_e(n) \right)$. \\

\subsection{Experimental determination of the sideband rate} \label{sec:SidebandRate}
%The sideband rate $\Omega_{SB,0}$ is estimated in the following way. The argument of the Bessel function in equation \label{eq:SBrate} is directly related to the frequency shift $\delta f_q$ of the qubit due to the dither drive. We measure this shift spectroscopically and use the relation : 
%\begin{equation}
%\frac{8 E_C n_g^{\textrm{dither}}}{E_J} = \sqrt{\left( 1 + \frac{\delta f_q}{f_q} \right)^2 -1}
%\end{equation}
%to obtain $8 E_C n_g^{\textrm{dither}}/E_J \approx 0.09$. We determine the Rabi rate $\Omega_R$ by using the sideband drive to perform an incoherent Rabi drive on resonance with the qubit, and using the fact that $\Omega_R \propto \sqrt{n_{\textrm{photon}}} \propto \frac{1}{\omega_{\textrm{drive}}-\omega_c}$ to compute $\Omega_R$ at the sideband transition frequency. We estimate $\Omega_R \approx 2\pi\times 37$ MHz for the sideband drive used in figure 3 of the main text. Driving at larger power seems to create trivial heating of the mechanical bath. Finally we obtain $\Omega_{SB,0} \approx 2\pi\times 55$ kHz. 

\begin{figure}[ht] \centering
\includegraphics[width=1.0\columnwidth]{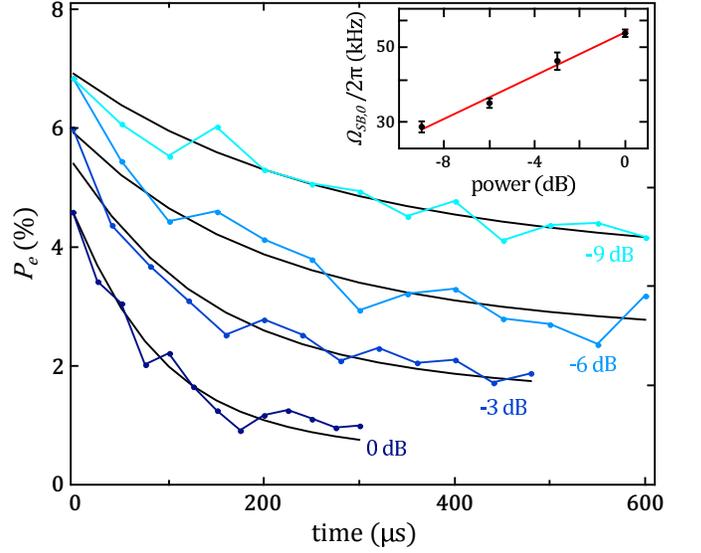}
\caption{ Qubit excited population (dots) as function of blue sideband duration (sideband drive centered around $n_\textrm{drive} \approx 1$) and measured at $n \approx 1$, i.e. for a spectroscopic tone at frequency $f \approx \left( \omega_q + 2 \chi_m \times 1 \right)/2\pi$. The four traces correspond to four different sideband drive powers (0 dB corresponds to -6 dBm at the top of the fridge, see Fig. \ref{SFig_Wiring}), and are offset for clarity. Solid lines are single parameter fit performed using lookup tables generated with our numerical simulation (equations \ref{eq:Master1} and \ref{eq:Master2}). Inset : extracted sideband rate $\Omega_{SB,0}$ (log scale, dots) and fit to a linear function (solid line, no offset at zero power). }
\label{SFig_SbRate}
\end{figure}

We estimate the sideband rate $\Omega_{SB,0}$ by measuring the dressed qubit excited probability evolution during a blue sideband drive. Fig. \ref{SFig_SbRate} shows the evolution of $P_e$, measured at $n\approx 1$, under a sideband drive centered around $n_\textrm{drive} \approx 1$, and at a few different sideband drive power. From our numerical simulations, we do not expect $P_e$ to be well described by an exponential decay. Nevertheless we observe a qualitative increase in the rate at which $P_e$ changes when the sideband drive power increases (and it is a proxy for the rate at which $P(n\approx 1)$ is depleted). In order to extract $\Omega_{SB,0}$, we generate lookup tables from our numerical model and perform a least square fit of these tables on the data. The only free parameter is $\Omega_{SB,0}$, and the results of the fits are plotted in the inset of Fig. \ref{SFig_SbRate}. Fitting with a linear function, we verify that $\Omega_{SB,0}$ is linear in the amplitude of the sideband drive. 

The data presented in Fig. 2 and 3 of the main text is taken with a sideband power corresponding to an estimated $\Omega_{SB,0} = 55.0 \pm 1.4$ kHz, which converts into $\Gamma_{SB}(n=0,n_{\textrm{drive}}=0) \approx 2\pi\times 3.3$ kHz.

\subsection{Discussion}
It is interesting to compare $\Gamma_{SB}(n=0,n_{\textrm{drive}}=0) \approx 2\pi\times 3.3$ kHz with $n_{th} \gamma_m \approx 2\pi\times 1.4$ kHz. Looking at these two numbers, the fact that we can move phonon populations efficiently might seem surprising. This effect comes from the fact that we have many levels coming into play. It can be understood semi-classically, by looking at rate equations between a few levels. Consider for instance a two-level system $\left\{ \ket{0},\ket{1} \right\}$, and assume it is coupled with a hot bath at a rate $\gamma$, so that $P(0) = P(1) = 1/2$ at thermal equilibrium. Turning on a drive takes populations from $\ket{0}$ to $\ket{1}$ at a rate $\Gamma = \gamma$ and the steady state populations will be $P(0) = 1/3$ and $P(1) = 2/3$. This is an inefficient drive as expected. However, the same drive will look more efficient already for 3 levels $\left\{ \ket{0},\ket{1}, \ket{2} \right\}$. At equilibrium with a hot bath, populations are $P(0) = P(1) = P(2) = 1/3$. With a drive taking populations from $\ket{0}$ to $\ket{1}$ and $\ket{1}$ to $\ket{2}$ at equal rates $\Gamma = \gamma$, the steady state populations are now $P(0) = 1/7$, $P(1) = 2/7$ and $P(2) = 4/7$. Similarly with 4 level, $P(0) = 1/15$, $P(1) = 2/15$, $P(2) = 4/15$ and $P(3) = 8/15$. So in the case of our mechanical oscillator under a sideband drive, many levels are being driven simultaneously and the imbalance between population of states that are far apart in the Fock space can be large even though $\Gamma_{SB}$ is comparable to $n_{th} \gamma_m$. Note that $\Gamma_{SB}(n,n_{\textrm{drive}})$ increases linearly with $n$, and the coupling rate between adjacent number states due to the hot bath scales the same way $\approx n n_{th} \gamma_m$, so the $n$ dependence cancels out. Generalizing this argument, we can estimate the expected occupation of the mechanical ground state under a red sideband. If we assume that our drive is not number sensitive, the final occupation should be on the order of $P(0) \approx \Gamma_{SB}/(\Gamma_{SB}+n_{th}\gamma_m)\approx 0.7$. We do not quite reach this occupation experimentally. In particular the actual number-sensitiveness of the sidebands would require us to either chirp our drive or wait for a long time (on the order of $1/\gamma_m$) for the higher number populations to relax all the way down close to the ground state.

\section{Mechanical damping rate measurement} \label{sec:gamma_m}
\begin{figure}[ht] \centering
\includegraphics[width=1.0\columnwidth]{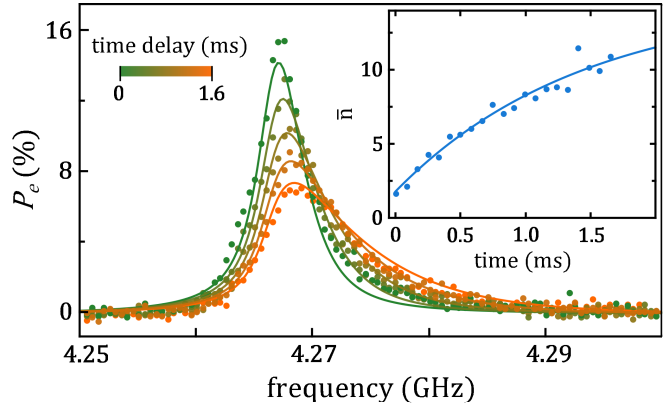}
\caption{ Thermalization of the mechanical oscillator to its environment, observed by measuring the dressed qubit spectroscopy at several different delays after a red sideband drive ($V_d = 6$ V). Dots are data and solid lines are fits assuming it is dressed by a mechanical thermal state. Inset : average mechanical occupancy $\bar{n}$ as a function of time after the red sideband. Dots are data and the solid line is a fit to an exponential with a characteristic damping time of 1.5 ms. }
\label{SFig_Thermalization}
\end{figure}

We measure the intrinsic damping rate of the mechanical oscillator by driving it out of equilibrium with a red or blue sideband and letting it relax to its thermal equilibrium. We record the average mechanical occupancy $\bar{n}$ during the relaxation process by measuring qubit spectroscopy and fitting or deconvolving this spectroscopy as explained in the main text. Fig. \ref{SFig_Thermalization} shows an example of such a measurement done after a red sideband drive. A few of the qubit spectroscopies taken with delays up to 1.6 ms delay are shown. When $\bar{n}$ is small (early times), a deconvolution would overestimate $\bar{n}$ because of the rectified added noise coming from enforcing positive probabilites. Therefore we make the approximation that the distribution is thermal and we perform a joined fit on all qubit spectroscopies at once. The extracted $\bar{n}$ fits well with a decaying exponential back to the thermal occupancy at equilibrium with the fridge. We do similar measurements after long blue sideband drives (not shown). Overall, we extract a mechanical decay time $1/\gamma_m = 1.72 \pm 0.32$ ms.

\section{Measurement setup and filtering} \label{sec:filtering}

\begin{figure*}[ht] \centering
\includegraphics[width=0.8\textwidth]{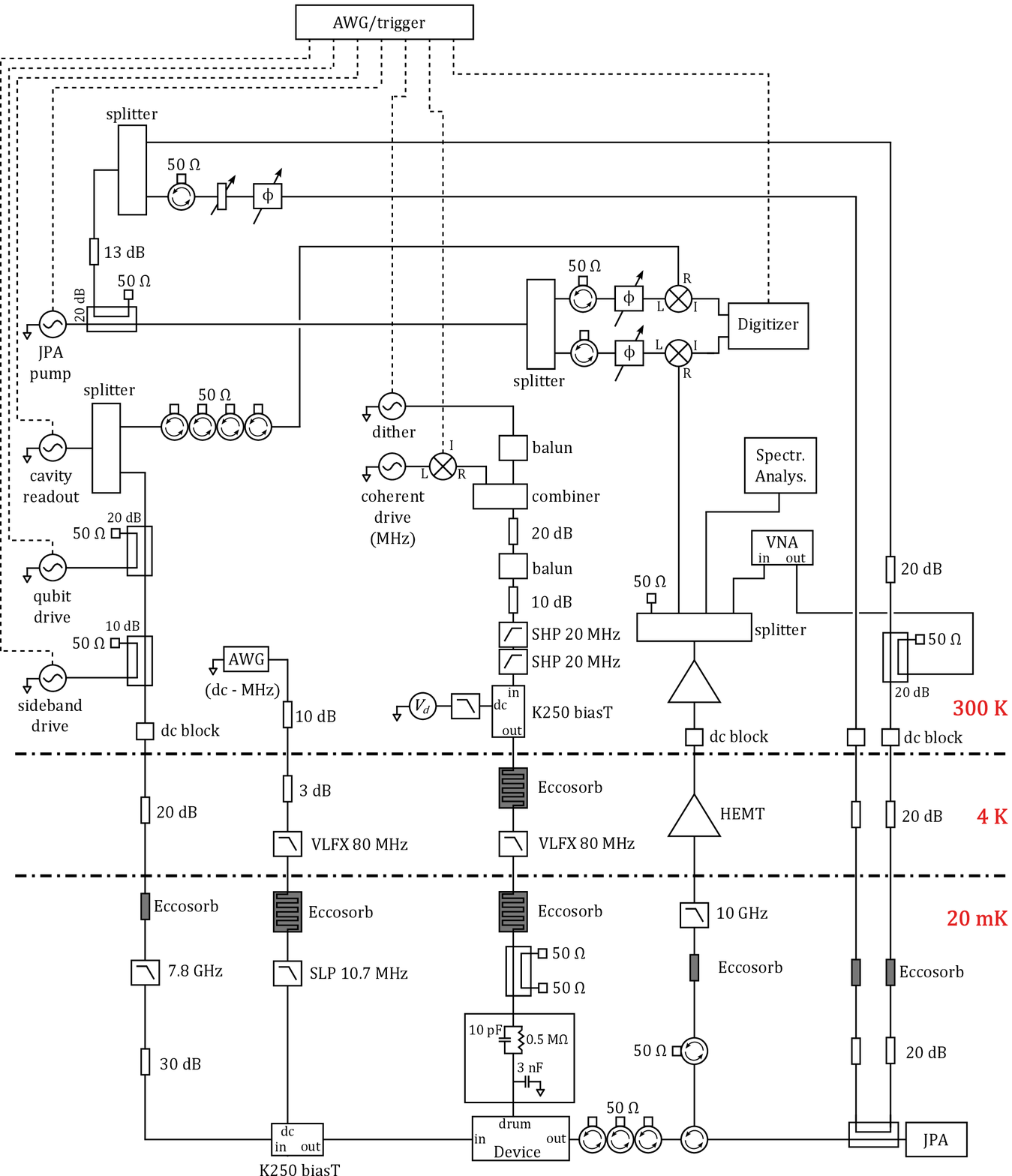}
\caption{Detailed wiring diagram.  }
\label{SFig_Wiring}
\end{figure*}

The full wiring diagram is given in Fig. \ref{SFig_Wiring}, showing room temperature and cryogenic setups. The filtering of the dc bias line is relatively demanding. It must provide a cold radiation environment (base temperature of our dilution refrigerator $\approx 13$ mK) to the mechanical oscillator at 25 MHz, while allowing weak rf drives at those frequencies. The home-made filter placed at base temperature is designed to achieve those goals. It behaves as a low-pass filter at low frequency with a 500 Hz cutoff. It becomes a voltage divider at 500 kHz with about 50 dB insertion loss up to about 60 MHz, ensuring cold radiation temperature at mechanical frequency. This filter thus behaves as a reflective attenuator, avoiding excessive heat load on the base temperature stage of the fridge for large drives at mechanical frequency. Seen at its output port, its impedance also has a small real part at rf frequencies to avoid damping of the mechanical oscillator (when $V_d \neq 0$, motion will dissipate current in any resistor in series and this will damp the motion). From the perspective of the qubit, around 4 GHz, this line should not look like a well coupled transmission line. We achieve this with a loss less reflective filter (two-stage LC filter) on chip, as shown in Fig. \ref{SFig_Device}(a) and (b). This filter has a cutoff frequency around 1 GHz and provides 52 dB of isolation at qubit frequency. We believe our qubit decay time $T_1 \approx 200$ ns to be limited by dielectric loss between the drum plates, not by decay into the dc bias line (which we estimate would yield $T_1 \approx 5$ $\mu$s). Finally, the entire filtering on the dc bias line should not add noise (in particular at low frequencies) and should support large voltages (design to support 30 V). \\

We add another dc voltage input $V_g$ with a cold bias Tee at the input of the cavity to allow fast control of the qubit gate voltage with a 10 MHz bandwidth. In practice, both $V_g$ and the drum voltage $V_d$ couple to the qubit gate charge but we keep $V_d$ fixed and use $V_g$ for fast calibration and control of the qubit gate charge $n_g$. \\

The microwave lines have standard attenuation and filtering. The output of the cavity goes to a Josephson parametric amplifier (JPA) which is pumped 4 MHz detuned from the cavity probe tone. Combined with JPA pump cancellation, this ensures that the residual pump power leaking in to the cavity is extremely small. We thereby perform a heterodyne detection. \\

%\bibliography{Library}

%

\end{document}